%% file: main.tex
\documentclass[acmsmall, screen]{acmart}

\AtBeginDocument{%
  }

\setcopyright{acmlicensed}
\copyrightyear{2025}
\acmYear{2025}
\acmDOI{XXXXXXX.XXXXXXX}

\acmJournal{CSUR}

\usepackage[utf8]{inputenc}
\usepackage[shortlabels]{enumitem}
\usepackage{url}
\usepackage{xspace}
\usepackage{comment}

\usepackage{amsmath}
\usepackage{amsfonts}
\usepackage{mathtools}
\usepackage{eucal}

\usepackage{graphicx}
\usepackage{subcaption}
\usepackage[export]{adjustbox}
\usepackage{forest}
\usepackage{wrapfig}
\usepackage{float}

\usepackage{tikz}
\usetikzlibrary{patterns,shapes,arrows,positioning,fit,backgrounds,calc}

\usepackage{tabularx}
\usepackage{multirow}
\usepackage{array}
\usepackage{booktabs}
\usepackage{colortbl}
\usepackage{arydshln}

\usepackage{xcolor}
\definecolor{darkblue}{rgb}{0.0, 0.0, 0.55}

\usepackage[normalem]{ulem}
\usepackage{textcomp}
\usepackage{soul}
\usepackage{pifont}
\usepackage{bbding}

\usepackage{fancybox}
\usepackage{tcolorbox}

\usepackage{algpseudocode}
\usepackage[linesnumbered,ruled,lined]{algorithm2e}

\usepackage{balance}
\usepackage{caption}

\usepackage{hyperref}
\usepackage{hyperxmp}

\begin{document}

\title{Large Language Model for Verilog Code Generation: Literature Review and the Road Ahead}

\author{Guang Yang}
\orcid{0000-0002-3374-6680}
\affiliation{
  \institution{Zhejiang University \& Northwestern Polytechnical University}
  \country{China}
}

\author{Wei Zheng}
\authornote{Corresponding Author: Wei Zheng <wzheng@nwpu.edu.cn>.}
\author{Dong Liang}
\author{Peng Hu}
\author{Yukui Yang}
\author{Shaohang Peng}
\author{Zhenghan Li}
\author{Jiahui Feng}
\author{Xiao Wei}
\author{Kexin Sun}
\author{Deyuan Ma}
\author{Haotian Cheng}
\author{Yiheng Shen}
\affiliation{%
    \institution{Northwestern Polytechnical University}
    \country{China}
}

\author{Xiang Chen}
\affiliation{%
\institution{Nantong University}
\country{China}
}

\author{Xing HU}
\authornote{Corresponding Author: Xing Hu <xinghu@zju.edu.cn>.}
\affiliation{%
\institution{Zhejiang University}
\country{China}
}

\author{Terry Yue Zhuo}
\affiliation{%
\institution{Monash University}
\country{Australia}
}

\author{David Lo}
\affiliation{%
\institution{Singapore Management University}
\country{Singapore}
}

\renewcommand{\shortauthors}{Yang et al.}

\begin{abstract}
Code generation represents a critical intersection of Software Engineering (SE) and Artificial Intelligence (AI). 
Within this broader landscape, Verilog, as a representative hardware description language (HDL), is fundamental to Electronic Design Automation (EDA), recent research has increasingly focused on leveraging Large Language Models (LLMs) to automate Verilog code generation, particularly at the Register Transfer Level (RTL) design.
Despite growing interest, a comprehensive survey of this domain remains absent.
This review addresses this gap by providing a systematic literature review of LLM-based Verilog code generation, analyzing 102 papers (70 published and 32 high-quality preprints) from SE, AI, and EDA venues. We structure our analysis around four key research questions: (1) identifying the LLMs utilized, (2) examining evaluation datasets and metrics, (3) categorizing generation techniques, and (4) analyzing alignment approaches. 
Furthermore, we synthesize findings to identify critical limitations in current studies regarding effectiveness and integration. 
Finally, we outline a roadmap highlighting potential opportunities for future research in LLM-assisted hardware design. 
\end{abstract}



\maketitle

\input{sec/01_introduction}
\input{sec/02_background}
\input{sec/03_methodology}
\input{sec/04_RQ1}
\input{sec/05_RQ2}
\input{sec/06_RQ3}
\input{sec/07_RQ4}
\input{sec/08_discussion}
\input{sec/09_conclusion}

\section{Acknowledgments}
The authors would like to thank the editors and the anonymous reviewers for their insightful comments and suggestions, which can substantially improve the quality of this work. 
This work was partially supported by the National Natural Science Foundation of China (NSFC, No. 62141208).
Guang Yang is also supported by the Postdoctoral Fellowship Program of CPSF under Grant Number GZC20260902.
  

\bibliographystyle{ACM-Reference-Format}
\bibliography{main}

\end{document}

%% file: sec/01_introduction.tex
\section{Introduction}


Large Language Models (LLMs) have revolutionized code generation, with models like Codex~\cite{1} and CodeLlama~\cite{2} demonstrating exceptional proficiency in general-purpose languages~\cite{3}. This success has rapidly extended to specialized domains, including scientific computing~\cite{4}, web development~\cite{5}, and domain-specific languages~\cite{6}.
However, hardware design presents distinct challenges compared to software development. 
The EDA industry faces widening productivity gaps due to escalating circuit complexity~\cite{8} and the specialized expertise required for HDLs like Verilog (e.g., timing and architecture). 
These factors, combined with labor-intensive manual workflows and severe talent shortages, create significant bottlenecks in modern hardware development.



Research on LLM-based Verilog generation has expanded dramatically, from a single study in 2020~\cite{9} to 66 papers in 2025 (see Figure~\ref{fig:paper_dictribution}). 
This research spans AI, EDA, and interdisciplinary venues, leading to fragmented knowledge with inconsistent terminologies, evaluation metrics, and optimization approaches.
Despite this growth, the lack of a systematic review hinders the community from unifying these contributions, establishing best practices, and identifying future directions.



To address this gap, we conducted a Systematic Literature Review (SLR)~\cite{10} using the Quasi-Gold Standard (QGS) strategy~\cite{11}. 
As shown in Figure~\ref{fig:process}, we combined manual searches of premier venues (e.g., AAAI, DAC) with automated searches across six major databases (e.g., IEEE Xplore, arXiv). 
Following a rigorous three-stage filtering and snowballing process, we selected 102 high-quality papers (70 peer-reviewed, 32 preprints) from 2020 to October 2025. 
This review answers four key research questions concerning the employed LLMs (RQ1), datasets and metrics (RQ2), optimization techniques (RQ3), and alignment approaches (RQ4).

\begin{table*}[!b]
  \centering
  \caption{Comparison with Related Literature Reviews}
  \label{tab:related_surveys}
  \resizebox{\textwidth}{!}{%
  \begin{tabular}{|l|c|c|c|c|}
  \hline
  \textbf{Survey} & \textbf{Year} & \textbf{Papers} & \textbf{Topics} & \textbf{Verilog Focus} \\
  \hline
  \multicolumn{5}{|c|}{\textbf{General Code Generation Surveys}} \\
  \hline
  Jiang et al.~\cite{12} & 2024 & 235 & General Programming Languages & \ding{55} \\
  Joel et al.~\cite{6} & 2024 & 111 & Low-Resource and Domain-Specific Programming Languages & Partial\\
  Chen et al.~\cite{13} & 2025 & 601 & Deep Learning-based Software Engineering & \ding{55} \\
  \hline
  \multicolumn{5}{|c|}{\textbf{EDA-focused Surveys}} \\
  \hline
  Fang et al.~\cite{14} & 2025 & 250 & Layout/Netlists/HDL/Assertion Generation & Partial \\
  He et al.~\cite{15} & 2024 & 211 & EDA Generation, Verification, and Debugging & Partial \\
  Chen et al.~\cite{16} & 2024 & $\approx 200$ & A Paradigm Shift from AI4EDA towards AI-rooted EDA (full-workflow AI4EDA) & Partial \\
  \hline
  \multicolumn{5}{|c|}{\textbf{Our Verilog-specific Survey}} \\
  \hline
  \textbf{Our work} & \textbf{2025} & \textbf{102} & Verilog Code Generation & \textbf{\checkmark} \\
  \hline
  \end{tabular}%
  }
  \end{table*}

\textbf{Related Literature Reviews.} 
While several surveys have examined LLM applications in code generation, software engineering, and hardware design automation, our work represents the first comprehensive review focusing specifically on Verilog code generation. 
Table~\ref{tab:related_surveys} summarizes how our work differs from existing literature reviews. 
Prior surveys fall into two primary categories: (1) \textit{\underline{general code generation surveys}} that predominantly concentrate on high-level programming languages like Python and Java but neglect hardware description languages, and (2) \textit{\underline{broad EDA surveys}} that cover multiple aspects of hardware design automation without in-depth analysis of LLM-based Verilog generation.

Existing surveys on general code generation~\cite{12,17,6} overlook hardware-specific challenges like timing and synthesizability. 
Similarly, broader EDA surveys~\cite{16,15,14} touch upon RTL generation but lack systematic analysis of Verilog-specific datasets and methodologies. 
In contrast, our survey provides the first comprehensive review dedicated to LLM-based Verilog generation, offering a systematic analysis of models, datasets, evaluation metrics, and alignment techniques from both AI and hardware perspectives.


\begin{figure*}[t]
  \centering 
  \begin{adjustbox}{width=1.0\columnwidth}
    \begin{forest}
for tree={
    rounded corners,
    child anchor=west,
    parent anchor=east,
    grow'=east,  
    text width=4cm,%
    draw=darkblue,
    anchor=west,
    node options={align=center},
    edge path={
      \noexpand\path[\forestoption{edge}]
        (.child anchor) -| +(-5pt,0) -- +(-5pt,0) |-
        (!u.parent anchor)\forestoption{edge label};
    },
    where n children=0{text width=7cm}{}
  },
  [LLM for Verilog Code Generation
    [Background and Preliminary \S\ref{sec:background}],
    [Review Methodology \S\ref{sec:method}],
    [LLMs Used for Verilog Code Generation \S\ref{RQ1}
      [Base LLMs \S\ref{RQ1:BaseLLMs}]
      [Instruction-Tuned LLMs \S\ref{RQ1:ITLLMs}]
    ],
    [Verilog-specific Datasets and Evaluation Metrics \S\ref{RQ2}
      [Dataset \S\ref{RQ2.1}
       [Benchmark \S\ref{RQ2.1.1}]
       [Instruction Tuning \S\ref{RQ2.1.2}]
      ]
      [Evaluation Metrics\S\ref{RQ2.2}
       [Static-based Metrics \S\ref{RQ2.2.1}]
       [Execution-based Metrics \S\ref{RQ2.2.2}]
       [LLM-based Metrics \S\ref{RQ2.2.3}]
      ]
    ],
    [Adaptation and Optimization Techniques \S\ref{RQ3}
      [Training-free Methods \S\ref{RQ3.1}
       [EDA-Tool Feedback \S\ref{RQ3.1.1}
        [Single-Agent Systems]
        [Multi-Agents Systems]
       ]
       [Prompt Engineering \S\ref{RQ3.1.2}]
       [Inference Optimization \S\ref{RQ3.1.3}]
      ]
      [Training-based Methods \S\ref{RQ3.2}
       [Pre-training \S\ref{RQ3.2.1}
        [Training from Scratch]
        [Continued Pre-training]
       ]
       [Supervised Fine-tuning \S\ref{RQ3.2.2}
        [Data-Centric Tuning]
        [Strategy-Centric Tuning]
        [Multi-task Tuning]
        [Knowledge-enhanced Tuning]
       ]
       [Reinforcement Learning \S\ref{RQ3.2.3}
        [Structure-based Rewards]
        [Tool-based Feedback]
        [Multi-objective Optimization]
       ]
      ]
    ],
    [Verilog-specific Alignment Approaches \S\ref{RQ4}
      [Security \S\ref{RQ4.1}]
      [Efficiency \S\ref{RQ4.2}]
      [Copyright \S\ref{RQ4.3}]
      [Hallucinations \S\ref{RQ4.4}]
    ],
    [The Road Ahead \S\ref{ChaAndOpp}
       [Cross-Study Evidence Synthesis \S\ref{synthesis}]
       [Limitations \S\ref{Challenges}],
       [Roadmap \S\ref{Roadmap}],
    ],
  ]
\end{forest}
\end{adjustbox}
\caption{Structure of This Survey} 
\label{fig:structure}
\end{figure*}
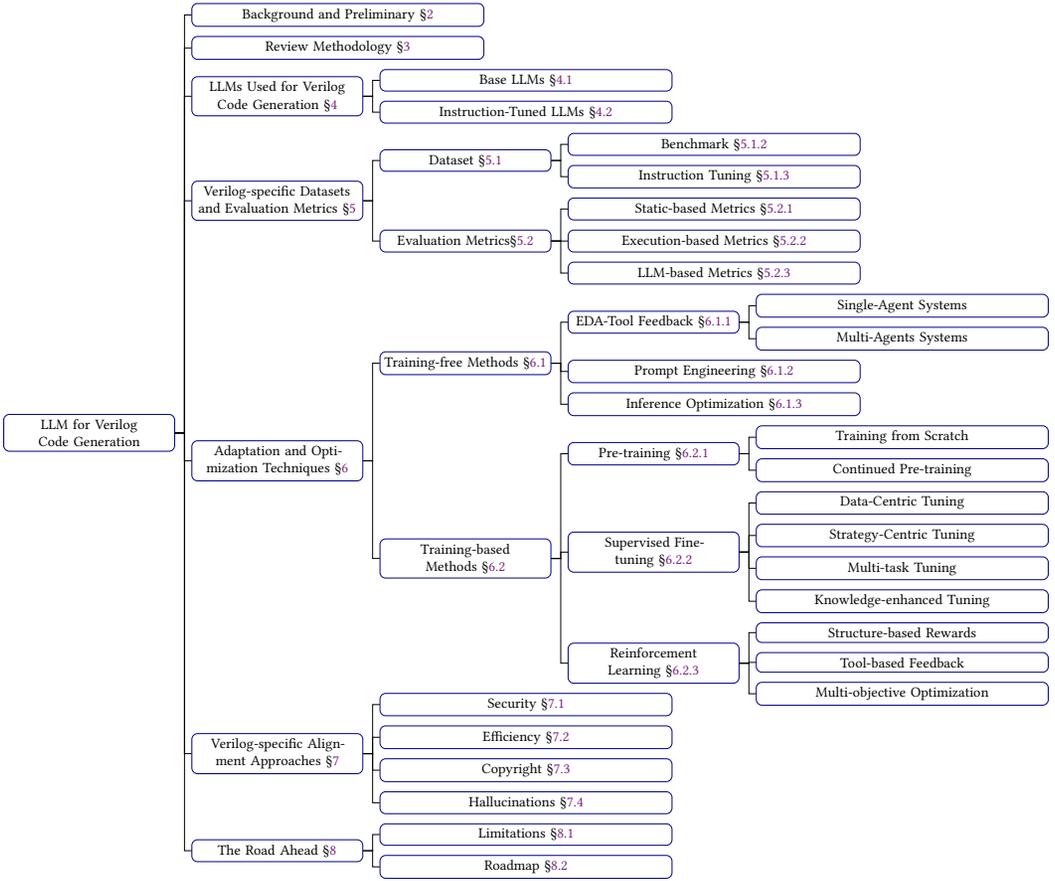






In general, this study makes the following \textbf{contributions}:

\begin{itemize}[leftmargin=1em]
  \item We present the first systematic literature review on LLM-based Verilog code generation, analyzing 102 papers (70 peer-reviewed, 32 high-quality preprints) spanning 2020-2025.

  \item We provide comprehensive taxonomy and trend analysis of LLMs for Verilog generation (RQ1), systematically analyze dataset construction and evaluation evolution across 28 benchmarks and 34 training datasets (RQ2), investigate adaptation and optimization techniques (RQ3), and examine alignment techniques addressing human-centric requirements including security, efficiency, copyright, and hallucinations (RQ4).

\item We identify key research limitations and propose a comprehensive roadmap for future directions in LLM-assisted hardware design.

\end{itemize}

\textbf{Survey Structure.} 
Figure~\ref{fig:structure} illustrates the organization of this survey. Sections~\ref{sec:background} and~\ref{sec:method} present the background and review methodology. The core analysis follows in Sections~\ref{RQ1} to~\ref{RQ4}, which respectively examine the used LLMs, datasets and metrics, generation techniques, and alignment strategies. Finally, Sections~\ref{ChaAndOpp} and~\ref{threats} discuss open challenges and future roadmaps, followed by the conclusion in Section~\ref{conclusion}.

%% file: sec/02_background.tex
\section{Background and Preliminaries}
~\label{sec:background}

\begin{figure*}[!t]
    \centering
    \includegraphics[width=0.8\linewidth]{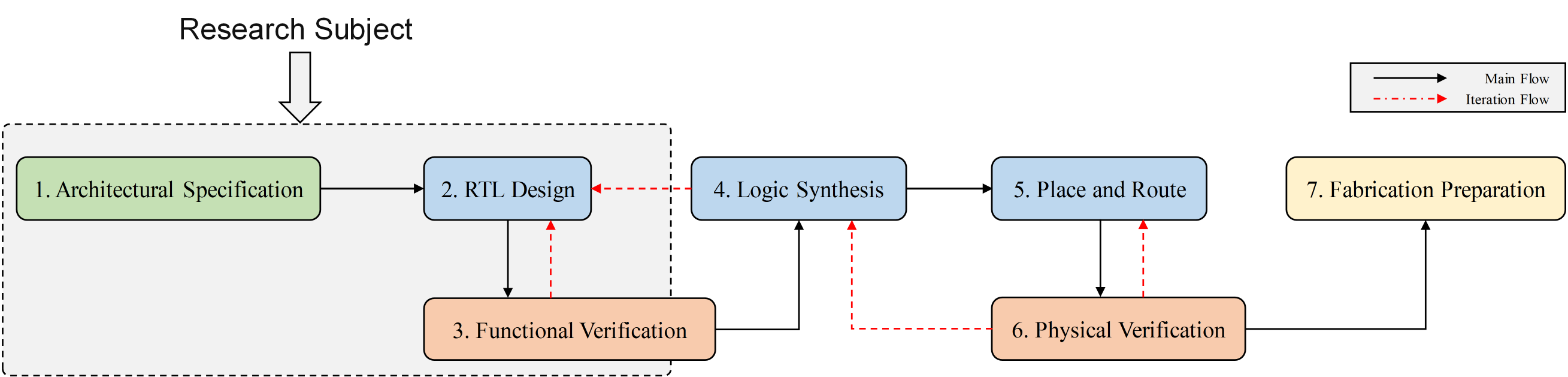}
    \caption{The EDA workflow for integrated circuit development.}
    \label{fig:workflow}
\end{figure*}


As shown in Figure~\ref{fig:workflow}, the EDA workflow begins with \textit{Architectural Specification} (Phase 1) and \textit{RTL Design} (Phase 2), where HDLs like Verilog describe hardware architecture. 
\textit{Functional Verification} (Phase 3) then validates correctness via testbenches, forming an iterative loop with RTL design. 
Subsequent phases include \textit{Logic Synthesis} (Phase 4), \textit{Place and Route} (Phase 5), \textit{Physical Verification} (Phase 6), and finally \textit{Fabrication Preparation} (Phase 7).

Verilog code generation (dashed box in Figure~\ref{fig:workflow}) primarily targets the first three phases. The quality of generated code is critical, directly impacting downstream synthesis, timing, and power metrics. This task differs fundamentally from software generation due to:
\begin{itemize}
  \item \textbf{Concurrency:} Hardware is inherently parallel, unlike sequential software.
  \item \textbf{Timing Constraints:} Designs must satisfy strict setup/hold times and delays.
  \item \textbf{Physical Limitations:} Code must respect area, power, and routing constraints.
  \item \textbf{Synthesizability:} Verification requires ensuring both functional and timing correctness.
  \item \textbf{Domain Expertise:} It demands deep knowledge of digital logic and architecture.
\end{itemize}
These characteristics highlight the unique challenges of applying LLMs to hardware design.

Formally, LLM-based Verilog generation is a mapping $f_{\theta}: (D, I) \rightarrow V$, where $\theta$ are model parameters, $D$ is the natural language specification (functional and constraint requirements), $I$ represents optional multimodal inputs (e.g., diagrams), and $V$ is the generated code. Most current approaches simplify this to text-only generation $f_{\theta}: D \rightarrow V$.

Correctness is evaluated via a function $verify: (V, S, T) \rightarrow \{pass, fail\} \times M$, where $S$ is the reference specification and $T$ denotes testbenches. This process typically includes: (1) \textit{syntactic validation} (compilability); (2) \textit{functional correctness} (simulation where $\forall t \in T: sim(V, t) = expected(t)$); (3) \textit{semantic similarity} (AST or text metrics); and (4) \textit{formal equivalence} (mathematical proof against reference designs).



%% file: sec/03_methodology.tex
\section{Review Methodology}
~\label{sec:method}

\begin{figure}[t]
    \centering
    \includegraphics[width=\linewidth]{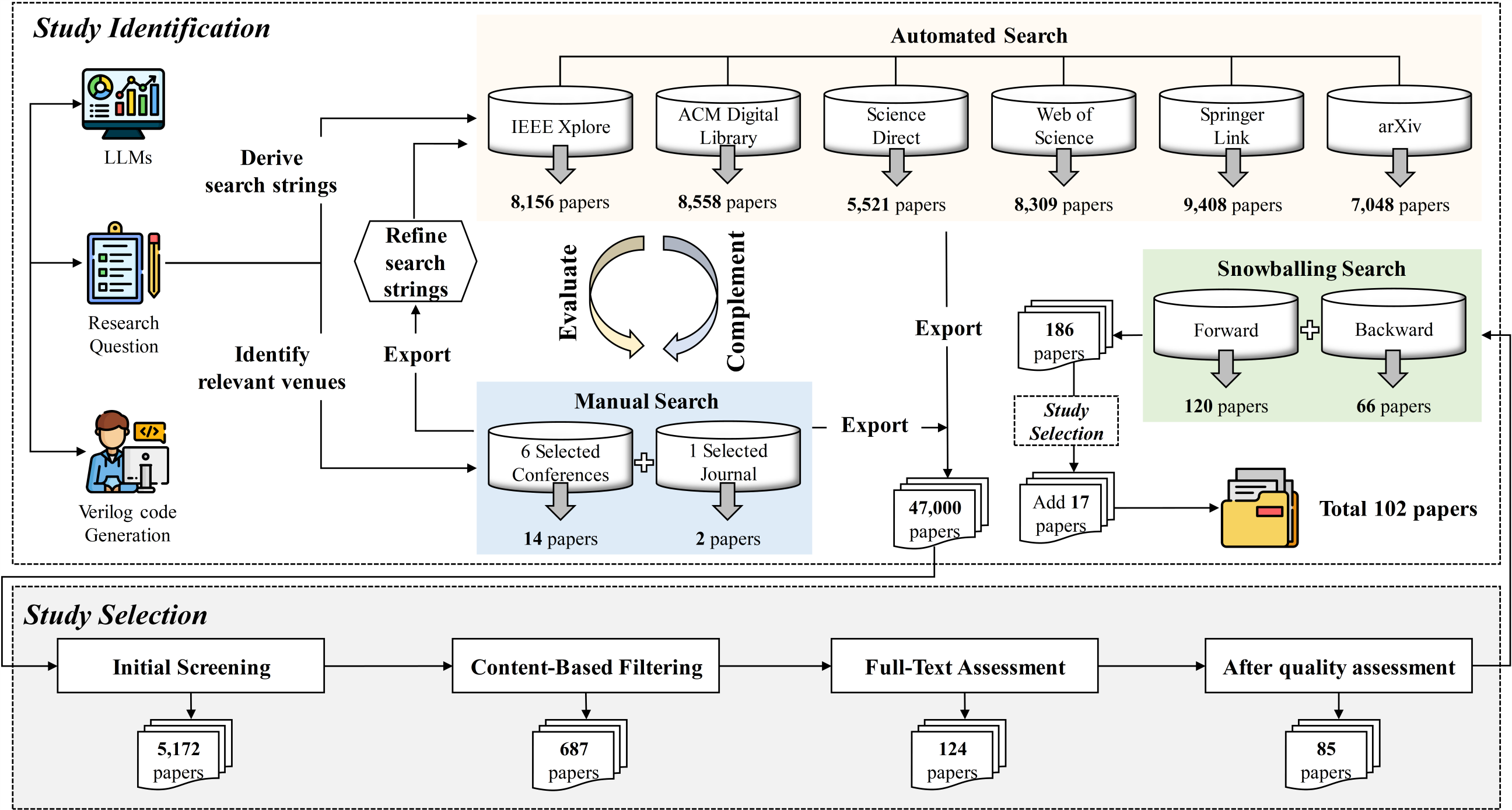}
    \caption{Systematic literature review process for study identification and selection. 
    }
    \label{fig:process}
\end{figure}

Following the guide by Kitchenham et al.~\cite{10}, which is used in most other SLRs~\cite{17, 24}, our methodology is organized as follows: planning the review (i.e., Section~\ref{sec:3_1}, \ref{sec:3_2}), conducting the review (i.e., Section~\ref{sec:3_3}), and analyzing the basic review results (i.e., Section~\ref{sec:3_4}).

\subsection{Research Question}
\label{sec:3_1}
This study thus aims to answer the following research questions:

\textbf{RQ1: What LLMs have been employed for Verilog code generation?} 
This question establishes a taxonomy of model architectures (Base vs. Instruction-Tuned LLMs, open vs. closed-source), revealing adoption trends and the evolution of model preferences in this domain.

\textbf{RQ2: How are Verilog-specific datasets constructed and evaluated?} 
We investigate dataset methodologies and evaluation metrics, which are critical for interpreting model performance, ensuring benchmark validity, and guiding future dataset optimization.

\textbf{RQ3: What adaptation and optimization techniques are applied?} 
This RQ examines strategies for customizing general-purpose LLMs to Verilog's unique constraints, identifying effective methods to enhance domain-specific performance.

\textbf{RQ4: What alignment techniques address human-centric requirements?} 
We analyze alignment approaches ensuring generated code conforms to human intent, addressing critical risks such as security vulnerabilities, copyright infringement, and trustworthiness.

\subsection{Search Strategy}
\label{sec:3_2}

As shown in Fig.\ref{fig:process}, we employed the ``Quasi-Gold Standard'' (QGS)~\cite{11} approach, combining manual identification of key studies with automated string-based searches and snowballing techniques. Rigorous filtering was then applied to select relevant papers. Our search targets publications from 2020 (the field's inception~\cite{9}) to 2025 to ensure comprehensive coverage.




\subsubsection{Search Items}

During the manual search phase, we selected eight premier conferences and journals (i.e., AAAI, ACL, ICML, ICLR, NeurIPS, DAC, MLCAD, and TCAD), and systematically retrieved papers applying LLMs to Verilog code generation tasks.
We identified 16 papers (14 papers from conferences and 2 papers from journals) that aligned with our research objectives. 
These 16 relevant publications constitute the foundational corpus for constructing our QGS, serving as the basis for extracting comprehensive search terms and validating our automated search strategy.
Our search string combines keywords for "Verilog code generation" and "LLMs", requiring matches from both categories to ensure precision.
The comprehensive search keyword sets are as follows:

\begin{itemize}
    \item \textbf{Keywords related to Verilog code generation (12 terms):} Verilog, HDL, Hardware Description Language, RTL, Register Transfer Level, Digital Design, Hardware Design, Electronic Design Automation, EDA, Verilog Generation, FPGA, ASIC
    \item \textbf{Keywords related to LLMs (13 terms):} LLM, Large Language Model, Language Model, GPT, ChatGPT, Transformer, fine-tuning, prompt engineering, In-context learning, Natural Language Processing, NLP, Machine Learning, AI
\end{itemize}



\subsubsection{Search Databases}
Following the establishment of our search strings, we conducted systematic automated searches across six widely recognized academic databases: 
IEEE Xplore~\cite{25}, ACM Digital Library~\cite{26}, ScienceDirect~\cite{27},
Web of Science~\cite{28}, SpringerLink~\cite{29}, and arXiv~\cite{30}.
Consistent with our temporal scope targeting publications from 2020 onward, we applied appropriate date filters to each database search.

\subsection{Search Selection}
\label{sec:3_3}
\begin{table}[t]
    \caption{Inclusion criteria and Exclusion criteria.}
    \resizebox{1\textwidth}{!}{
    \begin{tabular}{ll}
    \toprule
    \multicolumn{2}{l}{\textbf{Inclusion criteria}} \\ 
    \midrule
    1) & The paper claims that an LLM is used.  \\
    2) & The paper claims that the study involves to solve Verilog code generation. \\
    3) & The paper with accessible full text and must be written in English.    \\ 
    \midrule
    \multicolumn{2}{l}{\textbf{Exclusion criteria}}   \\ 
    \midrule
    1) & Short papers whose number of pages is less than 5.\\
    2) & Books, records, theses, tool demos papers, editorials, or venues not subject to a full peer-review process.\\
    3) & The paper is a literature review or survey. \\
    4) & The paper mentions LLMs only in future work or discussions rather than using LLMs in the approach.\\
    5) & The paper does not involve Verilog code generation. \\
    6) & Duplicate papers or similar studies authored by the same authors.\\
    \bottomrule
    \end{tabular}
    }
    \label{tab:criteria}
\end{table}

\subsubsection{Inclusion and Exclusion Criteria}
To ensure methodological rigor, we established inclusion and exclusion criteria (Table~\ref{tab:criteria}) based on established SLR practices~\cite{31}. 
These criteria ensure selected studies directly address our research questions.\footnote{For preprint papers released on arXiv, we performed manual quality assessments to determine their eligibility for inclusion based on methodological soundness and contribution relevance.}


Our paper selection process employed a three-stage filtering approach to systematically reduce the candidate pool:
\textcircled{1} \textbf{Stage 1 (Initial Screening):} Exclusion of short papers (< 5 pages) and duplicates (criteria 1, 6) reduced the corpus to 5,172 papers.
\textcircled{2} 
\textbf{Stage 2 (Content-Based Filtering):} Manual screening of venues, titles, and abstracts narrowed selection to 687 papers. We retained high-quality preprints but excluded non-peer-reviewed materials like theses and reports (criteria 2-3).
\textcircled{3} 
\textbf{Stage 3 (Full-Text Assessment):} Full-text review excluded studies focused solely on repair/testing without generation (criterion 5) or lacking implementation (criterion 4). This yielded 124 primary studies.




\subsubsection{Quality Assessment}

\begin{table}[t]
    \caption{Checklist of Quality Assessment Criteria (QAC).}
    \resizebox{1\textwidth}{!}{
    \begin{tabular}{rl}
    \toprule
    \textbf{ID} & \textbf{Quality Assessment Criteria} \\ 
    \midrule
    QAC1  & Was the study published in a prestigious (e.g., CCF ranking) venue?      \\
    QAC2  & Does the study make a contribution to the academic or industrial community? \\
    QAC3  & Does the study provide a clear description of the workflow and implementation of the proposed approach?   \\
    QAC4  & Are the experiment details, including datasets, baselines, and evaluation metrics, clearly outlined?      \\
    QAC5  & Do the findings from the experiments strongly support the main arguments presented in the study?          \\ 
    \bottomrule 
    \end{tabular}}
    \label{tab:assessment}
\end{table}

To ensure study quality and transparency, we implemented a rigorous assessment framework comprising five Quality Assessment Criteria (QACs) (Table~\ref{tab:assessment})~\cite{33}. 
Each criterion was evaluated on a 0-3 scale, with a total inclusion threshold set at 12 points (80\% of the maximum). 
For preprints lacking venue scores (i.e., QAC1 is 0), they were retained only if they demonstrated exceptional merit in other dimensions to meet the threshold. 

This process yielded a final corpus of 85 studies, consisting of 55 peer-reviewed publications and 30 high-quality preprints, balancing scholarly rigor with the inclusion of emerging research.




\subsubsection{Forward and Backward Snowballing}
To ensure comprehensive coverage, we conducted systematic forward and backward snowballing~\cite{34}. This supplementary strategy involves examining the reference lists of our primary studies (backward snowballing) to find missed relevant works, and utilizing citation databases to identify subsequent publications that cite our selected studies (forward snowballing), thereby capturing emerging developments.

The snowballing process initially yielded 186 additional candidates. These papers were subjected to our established selection pipeline, including screening against inclusion/exclusion criteria and quality assessment. Following this rigorous evaluation, we identified 15 additional relevant studies. This resulted in a final corpus of 102 papers (70 peer-reviewed publications and 32 high-quality preprints), which forms the empirical foundation of our review.


\subsection{Data Extraction and Analysis}
\label{sec:3_4}

\begin{figure}[t]
    \centering
    \subfloat[Distribution of papers across venues.] {\includegraphics[width=2.5in]{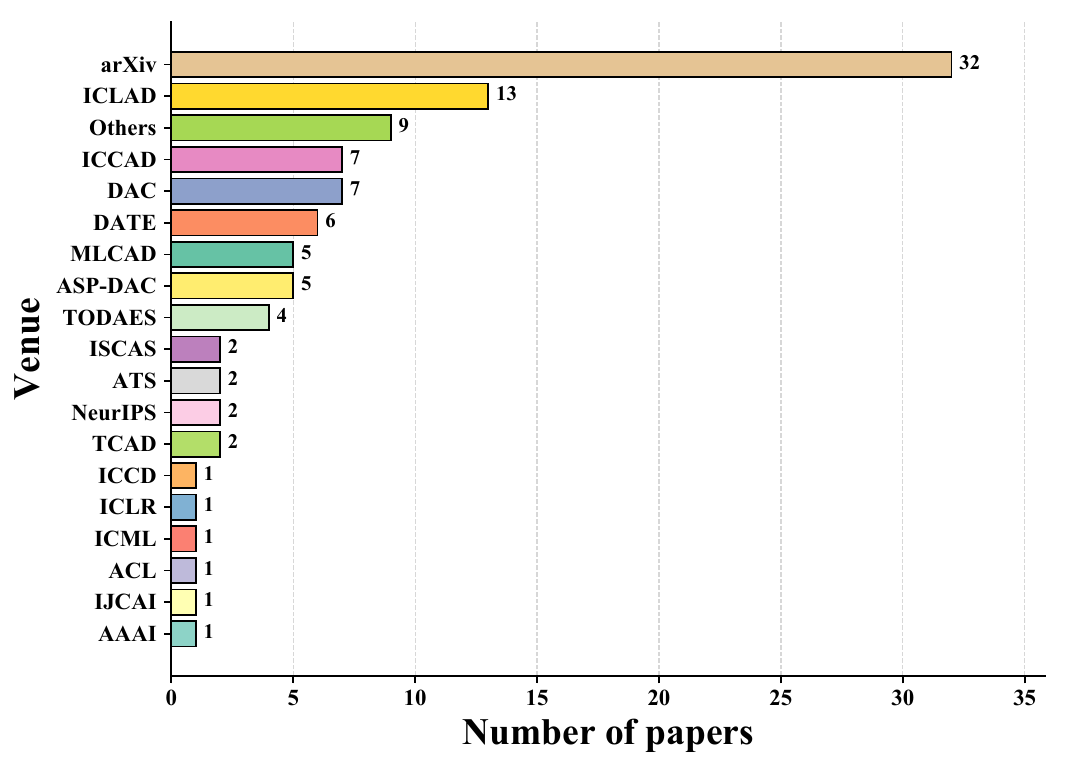}}
    \subfloat[Distribution of papers over years.]{\includegraphics[width=2.5in]{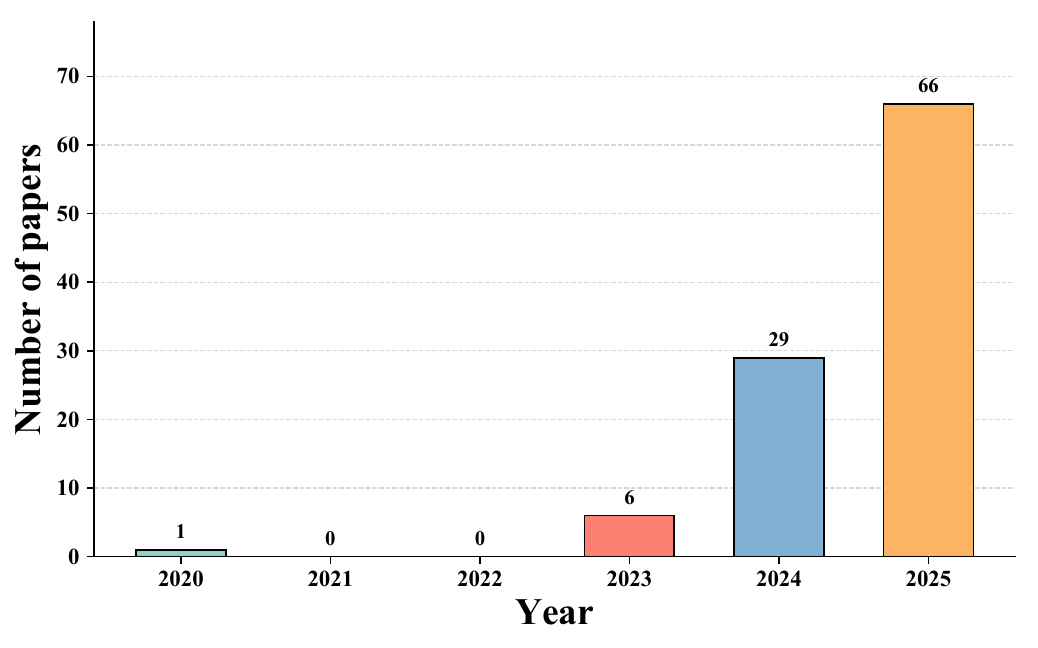}}
    \caption{Overview of the selected 102 papers' distribution.}
    \label{fig:paper_dictribution}
\end{figure}


Figure~\ref{fig:paper_dictribution} presents the distribution of the 102 selected papers and the temporal growth. 
From a single study in 2020 and a dormant period (2021-2022), publications surged to 6 in 2023, 29 in 2024, and 66 by September 2025. This exponential trajectory underscores the intensifying research interest in this domain.

%% file: sec/04_RQ1.tex
\section{RQ1: LLMs Employed for Verilog Code Generation}\label{RQ1}

\begin{table}[t]
    \centering
    \small 
    \caption{Usage of Open and Closed-Source LLM Families in Verilog Code Generation Studies (2023--2025)}
    \label{tab:llm_usage_all}
    \begin{tabular}{llcccc}
      \toprule
      \textbf{Type} & \textbf{LLM Family} & \textbf{2023} & \textbf{2024} & \textbf{2025} & \textbf{Total} \\
      \midrule
      Open-source  & Llama series      & 0 & 15 & 49  & 64  \\
      Open-source  & DeepSeek series   & 0 & 6  & 48  & 54  \\
      Open-source  & Qwen series       & 0 & 3  & 25  & 28  \\
      Open-source  & CodeGen series    & 3 & 6  & 3   & 12  \\
      Open-source  & StarCoder series  & 0 & 5  & 6   & 11  \\
      Open-source  & Others            & 2 & 11 & 22  & 35  \\
      \midrule
      \textbf{Open-source} & \textbf{Subtotal} & \textbf{5} & \textbf{46} & \textbf{153} & \textbf{204} \\
      \midrule
      Closed-source & GPT series       & 7 & 42 & 100 & 149 \\
      Closed-source & Claude series    & 0 & 5  & 14  & 19  \\
      Closed-source & Gemini           & 0 & 1  & 4   & 5   \\
      Closed-source & Others           & 0 & 3  & 3   & 6   \\
      \midrule
      \textbf{Closed-source} & \textbf{Subtotal} & \textbf{7} & \textbf{51} & \textbf{121} & \textbf{179} \\
      \midrule
      \multicolumn{2}{l}{\textbf{Overall}} & \textbf{12} & \textbf{97} & \textbf{274} & \textbf{383} \\
      \bottomrule
    \end{tabular}
  \end{table}

Through our systematic literature review of 102 included studies, we identify that researchers have employed a diverse range of LLMs for Verilog code generation tasks. 

\textbf{Statistical Scope.} Our analysis encompasses all LLMs utilized in the reviewed papers, whether for \textit{method implementation} or as \textit{baseline comparisons}, ensuring comprehensive coverage of the LLM adoption landscape in Verilog code generation research.
Based on their adaptation for Verilog tasks, we categorize these LLMs into two main types: \textbf{Base LLMs} (\textit{general-purpose models without Verilog-specific adaptation}) and \textbf{IT LLMs} (\textit{\underline{I}nstruction-\underline{T}uned models with Verilog-specific adaptation}).

\subsection{Analysis of Base LLMs}
\label{RQ1:BaseLLMs}
As summarized in Table~\ref{tab:llm_usage_all}, Base LLMs account for 383 total usages across all surveyed studies and thus constitute the dominant paradigm for Verilog code generation. 
Although these models are not explicitly designed for hardware description languages, their strong general-purpose programming capabilities already support competitive Verilog understanding and generation.

\subsubsection{Open-Source Base LLMs}
Open-source Base LLMs exhibit substantial diversity and adoption, reaching 204 usages (53.3\% of all Base LLM deployments). 
The landscape is structured around several major families:

\paragraph{(1) Llama Series} 
The Llama ecosystem~\cite{35} is the most widely adopted open-source family, with 64 total mentions. 
This reflects a clear preference for large, high-quality foundation models (e.g., CodeLlama, Llama3.x~\cite{2,36}) that provide strong general code-generation capabilities rather than Verilog-specific specialization.

\paragraph{(2) DeepSeek Series} 
DeepSeek models~\cite{37} achieve 54 mentions, ranking second among open-source families. 
The use of both code-centric and reasoning-oriented variants (e.g., DeepSeek-Coder, DeepSeek-R1~\cite{38,39}) suggests growing interest in leveraging advanced reasoning for complex Verilog design and debugging.

\paragraph{(3) Qwen Series} 
The Qwen family accumulates 28 mentions, emerging as a competitive open-source alternative. 
Models such as CodeQwen and Qwen2.5-Coder~\cite{40,41} strengthen code understanding and multi-language support, which is beneficial for heterogeneous hardware design toolchains.

\paragraph{(4) Others} 
Traditional code-oriented families such as CodeGen (12 mentions)~\cite{42,43} and StarCoder (11 mentions)~\cite{44,45} remain in use but are increasingly overshadowed by newer foundation families with larger scales and more diverse training data, indicating that general capability and ecosystem maturity often outweigh narrowly targeted code training.
A substantial portion (35 mentions) corresponds to various other open-source models (e.g., CodeGemma~\cite{46}, CodeT5+~\cite{47}, Phi~\cite{48,49}), 
highlighting the experimental nature of the field and researchers' willingness to explore alternative architectures and training paradigms.

\subsubsection{Closed-Source Base LLMs}
Closed-source Base LLMs also exhibit strong adoption, with 179 usages (46.7\% of all Base LLM deployments), predominantly provided by major technology companies.

\paragraph{(1) GPT Series} 
OpenAI's GPT series~\cite{51} dominates the closed-source segment with 149 mentions (83.2\% of closed-source usage). 
Different generations (e.g., GPT-3.5, GPT-4, GPT-4o) are widely employed for Verilog code generation, while newer reasoning-oriented models (such as GPT-o1 and GPT-o3) are increasingly explored for more challenging design and verification tasks.

\paragraph{(2) Other Models} 
Anthropic's Claude family~\cite{52} reaches 19 mentions, positioning itself as a strong competitor in closed-source Verilog applications. 
Google's Gemini~\cite{53} appears as a more recent entrant with 5 mentions, indicating emerging interest but still limited deployment compared to GPT-series models.

\subsubsection{Trends Analysis}

The field exhibits rapid expansion, with total Base LLM usages growing from 12 in 2023 to 274 in 2025 (\(2{,}183\%\) growth). 
This trajectory reflects both the fast-paced evolution of LLM technology and the community's increasing confidence in applying these models to hardware design.

We first analyze adoption trends from the perspective of Open-Source vs. Closed-Source models and then discuss architectural preferences:

\paragraph{(1) Open-Source vs. Closed-Source} 
Both categories grow rapidly, but with different trajectories and ecosystem dynamics:

\begin{itemize}
    \item \textbf{Open-Source LLMs:} Usages increase from 5 to 46 to 153 mentions, corresponding to a 233\% growth from 2024 to 2025. 
    The evolution shows a shift from traditional code models to large foundation families:
      \begin{itemize}
        \item \textbf{2023:} Dominated by classic code generators (e.g., CodeGen with 3 mentions).
        \item \textbf{2024:} Transition phase where foundation families (Llama: 15, DeepSeek: 6, Qwen: 3 mentions) begin to dominate.
        \item \textbf{2025:} Widespread adoption of the latest foundation models (Llama: 49, DeepSeek: 48, Qwen: 25 mentions), which become the de-facto standard for open-source Verilog generation.
      \end{itemize}
    \item \textbf{Closed-Source LLMs:} Usages grow from 7 to 51 to 121 mentions (137\% growth from 2024 to 2025), reflecting both consolidation and diversification:
      \begin{itemize}
        \item \textbf{2023:} Exclusively OpenAI models (GPT-3.5 and GPT-4, 7 mentions in total).
        \item \textbf{2024:} GPT remains dominant (42 mentions), while competitors such as Claude (5 mentions) and Gemini (1 mention) begin to appear.
        \item \textbf{2025:} Market expansion with new reasoning models (GPT-o1/o3) and stronger competition from Claude 3.5 and Gemini, although GPT still maintains a clear lead.
      \end{itemize}
  \end{itemize}

\paragraph{(2) Architectural Preferences} 
Most employed Base LLMs adopt decoder-only architectures, aligning with the mainstream autoregressive generation paradigm. 
A subset of studies explores encoder–decoder architectures (e.g., CodeT5+~\cite{47}), indicating that architectural diversity remains a promising research direction. 
In addition, vision–language models (e.g., GPT-4V, LlaVa~\cite{54}) are increasingly leveraged to translate visual design specifications or waveform diagrams into Verilog, pointing toward a multimodal future for hardware-oriented LLMs.

\begin{table*}[!t]
    \centering
    \caption{Instruction-Tuned LLMs for Verilog Code Generation (Classified by Model Weight Availability). }
    \label{tab:it_llms}
    \resizebox{\textwidth}{!}{%
    \begin{tabular}{|l|l|l|l|}
    \hline
    \textbf{Verilog-Specific LLM} & \textbf{Foundation Model} & \textbf{Weight} & \textbf{URL} \\
    \hline
    \multicolumn{4}{|l|}{\textbf{Closed-Source (Model Weights)} } \\
    \hline
    AutoVCoder~\cite{55} & CL/CQ/DS-Coder & Closed & -\\
    BetterV~\cite{56} & CL/CQ/DS-Coder & Closed & - \\
    CodeGen-Verilog~\cite{57} & CodeGen & Closed & - \\
    CraftRTL~\cite{58} & CL/DS-Coder/StarCoder2 & Closed & - \\
    DeepRTL~\cite{59} & CodeT5+ & Closed & - \\
    DeepRTL2~\cite{60} & Llama3.1/DS-Coder & Closed & - \\
    FreeV~\cite{61} & Llama3.1 & Closed & - \\
    ITERTL~\cite{62} & DS-Coder/DS-Coder-v2 & Closed  & -\\
    MEV-LLM~\cite{63} & CodeGen/Gemma & Closed & - \\
    OpenRTLSet~\cite{64} & Qwen2.5 & Closed & - \\
    PyraNet~\cite{65} & CL/DS-Coder & Closed & - \\
    RTL++~\cite{66} & CL & Closed & - \\
    RTLRepoCoder~\cite{67} & DS-Coder & Closed & - \\
    ScaleRTL~\cite{68} & DS-R1-Distill-Qwen & Closed & - \\
    Veritas~\cite{69} & Llama-3.2 & Closed  & -\\
    \hline
    \multicolumn{4}{|l|}{\textbf{Open-Source (Model Weights)} } \\
    \hline
    DAVE~\cite{9} & GPT-2 & Open & \url{https://shorturl.asia/VWPTn}\\
    ChipGPT~\cite{70} & Llama2/Llama3 & Open & \url{https://www.modelscope.cn/profile/changkaiyan}\\
    CodeV~\cite{71} & CQ/DS-Coder/QC & Open & \url{https://huggingface.co/zhuyaoyu}\\
    CodeV-R1~\cite{72} & QC & Open & \url{https://huggingface.co/zhuyaoyu}\\
    GEMMV~\cite{73} & DS-R1-Distill-Qwen/Llama3.1 & Open & \url{https://huggingface.co/bxsk2024}\\
    Hardware Phi-1.5B~\cite{74} & Phi-1.5B & Open & \url{https://huggingface.co/KSU-HW-SEC/Hardware_Phi_30k_version}\\
    HAVEN~\cite{75} & CL/DS-Coder/CQ & Open & \url{https://huggingface.co/yangyiyao}\\
    MG-Verilog~\cite{76} & CL & Open & \url{https://shorturl.asia/rDNSm}\\
    Mistral-Verilog~\cite{77} & Mistral & Open & \url{https://huggingface.co/emilgoh/mistral-verilog}\\
    Origen~\cite{78} & DS-Coder & Open & \url{https://huggingface.co/henryen/OriGen}\\
    ReasoningV~\cite{79} & QC & Open & \url{https://modelscope.cn/models/GipsyAI/ReasoningV-7B}\\
    RTLCoder~\cite{80} & Mistral/DS-Coder & Open & \url{https://github.com/hkust-zhiyao/RTL-Coder}\\
    VeriCoder~\cite{81} & QC & Open & \url{https://huggingface.co/LLM4Code/VeriCoder_Qwen14B}\\
    VeriGen~\cite{82} & CodeGen & Open & \url{https://huggingface.co/shailja}\\
    VeriLogos~\cite{83} & DS-Coder & Open & \url{https://huggingface.co/97kjmin/VeriLogos}\\
    VeriPrefer~\cite{84} & Mistral/CL/DS-Coder/CQ/QC & Open & \url{https://shorturl.asia/sMj3J}\\
    VeriReason~\cite{85} & QC/CL & Open & \url{https://shorturl.asia/yaNkf}\\
    VeriSeek~\cite{86} & DS-Coder & Open & \url{https://huggingface.co/LLM-EDA/VeriSeek}\\
    VeriThoughts~\cite{87} & QC & Open & \url{https://shorturl.asia/81f2a}\\
    \hline
    \end{tabular}%
    }
\footnotesize{CL: CodeLlama, DS: DeepSeek, CQ: CodeQwen, QC: Qwen2.5-Coder}
\end{table*}

\subsection{Analysis of Instruction Tuned LLMs}
\label{RQ1:ITLLMs}

As shown in Table~\ref{tab:it_llms}, we identify 34 instruction-tuned (IT) LLMs specifically built for Verilog code generation. 
A notable characteristic of this ecosystem is its relatively high degree of openness: 19 models (55.9\%) release public weights, while 15 (44.1\%) remain closed.

\subsubsection{Foundation Model}
Foundation model choices reveal a clear preference for code-specialized bases. 
Overall, 28 out of 34 IT LLMs (82.4\%) adopt coding-oriented foundations, with three major clusters:
DeepSeek-Coder variants (11 models, 32.4\%), Qwen coder series (9 models, 26.5\%), and CodeLlama derivatives (8 models, 23.5\%). 
The remaining 6 models (17.6\%) build on general-purpose foundations, led by Llama-family models and compact architectures such as GPT-2 and Phi-1.5B, which target more efficient deployment. 
Furthermore, 7 projects (20.6\%) experiment with multiple foundation models, suggesting that systematic comparison across bases is still an active research practice.



\subsubsection{Trends and Model Selection Insights}

Overall, IT LLM development for Verilog is consolidating around code-specialized foundations (DeepSeek-Coder, Qwen coder series, CodeLlama) and open-weight releases, which together facilitate reproducibility and fair comparison across studies. 
At the same time, there is a pronounced shift toward explicit reasoning capabilities (e.g., ``R1-style'' and ``Reasoning/Thoughts'' variants) to better support long-horizon design and verification tasks.

From a practical perspective, resource-constrained academic environments benefit most from open-source foundations such as DeepSeek-Coder, Llama3.1, and Qwen2.5-Coder, which balance performance, cost, and customizability, albeit at the expense of local compute requirements. 
For performance-critical applications, state-of-the-art closed-source models (notably GPT and Claude variants) typically deliver the strongest generation and reasoning quality, but incur higher monetary and API-dependency costs. 
For domain-specific Verilog research, we recommend prioritizing the 19 open-weight IT LLMs as starting points, with particular attention to reasoning-enhanced variants (e.g., CodeV-R1~\cite{72}, ReasoningV~\cite{79}, VeriReason~\cite{85}, VeriThoughts~\cite{87}), which are better aligned with the complex logical inference demands of modern hardware design.

\begin{tcolorbox}[left=4pt,right=4pt,top=2pt,bottom=2pt,boxrule=0.5pt]
    \textbf{Answer to RQ1:} 
We classify LLMs into two types: Base LLMs and Instruction Tuned (IT) LLMs, where base LLMs include both open-source models and closed-source models. 
Additionally, 34 domain-specific IT LLMs have been developed, with 19 providing open weights and the majority built upon code-specialized foundation models like DeepSeek-Coder and Qwen2.5-Coder.
\end{tcolorbox}

%% file: sec/05_RQ2.tex
\section{RQ2: Dataset and Evaluation Metrics for Verilog Code Generation}\label{RQ2}

\subsection{Dataset Construction}\label{RQ2.1}

We categorize datasets for Verilog code generation into two complementary families: (1) \textit{Benchmark datasets} for standardized evaluation, and (2) \textit{Instruct-tuning datasets} for supervised fine-tuning and reasoning augmentation. 

\subsubsection{Overview}
Tables~\ref{tab:Benchmark} and~\ref{tab:it_datasets} provide a comprehensive overview of the Verilog code generation dataset landscape. 
In terms of benchmarking, our analysis reveals 19 open-source benchmarks compared to 9 closed-source alternatives (2023--2025).
Similarly, in the instruction-tuning domain, open datasets (22 entries) also outnumber their closed-source items (12 entries).



\begin{table*}[!t]
    \centering
    \caption{Benchmark Datasets for Verilog Code Generation (Classified by Availability). }
    \label{tab:Benchmark}
    \resizebox{\textwidth}{!}{%
    \begin{tabular}{|l|l|l|l|l|l|l|}
    \hline
    \textbf{Dataset} & \textbf{Year} & \textbf{Input} & \textbf{Samples} & \textbf{Testbench} & \textbf{Available} & \text{URL}\\
    \hline
    \multicolumn{7}{|l|}{\textbf{Closed-Source} } \\
    \hline
    DAVE\_test~\cite{9} & 2020 & Text & 250 & No & Closed & - \\
    LLM-aided~\cite{89} & 2024 & Text & 10 & Yes & Closed & - \\
    MCTS~\cite{90} & 2024 & Text & 15 & Yes & Closed & - \\
    NL2Verilog~\cite{91} & 2024 & Text & 8 & Yes & Closed & - \\
    VGV~\cite{92} & 2024 & Text+Image & 20 & Yes & Closed & - \\
    AutoSilicon\_test~\cite{93} & 2025 & Text & 9 & Yes & Closed & - \\
    GEMMV\_test~\cite{73} & 2025 & Text & 10 & Yes & Closed & - \\
    HiVeGen\_test~\cite{94} & 2025 & Text & 4 & Yes & Closed & - \\
    ModelEval\_test~\cite{95} & 2025 & Text & 94 & Yes & Closed & - \\
    \hline
    \multicolumn{7}{|l|}{\textbf{Open-Source} } \\
    \hline
    ChipGPT~\cite{70} & 2023 & Text & 8 & Yes & Open & \url{https://zenodo.org/records/7953725} \\
    VeriGen\_test~\cite{88} & 2023 & Text & 17 & Yes & Open & \url{https://github.com/shailja-thakur/VGen}\\
    VerilogEval-v1~\cite{57} & 2023 & Text & 156/143 & Yes & Open & \url{https://github.com/NVlabs/verilog-eval/tree/release/1.0.0}\\
    AutoChip~\cite{96} & 2024 & Text & 138 & Yes & Open & \url{https://zenodo.org/records/10160723}\\
    ChipGPTV~\cite{97} & 2024 & Text+Image & 30 & Yes & Open & \url{https://github.com/aichipdesign/chipgptv}\\
    CreativEval~\cite{98} & 2024 & Text & 120 & Yes & Open & \url{https://github.com/matthewdelorenzo/CreativEval}\\
    Evaluatie\_LLMs~\cite{99} & 2024 & Text & 8 & Yes & Open & \url{https://zenodo.org/records/10947127}\\
    Hierarchical~\cite{100} & 2024 & Text & 10 & Yes & Open & \url{https://github.com/ajn313/ROME-LLM}\\
    PyHDL-Eval~\cite{154} & 2024 & Text & 168 & Yes & Open & \url{https://github.com/cornell-brg/pyhdl-eval}\\
    RTLLM-v1~\cite{101} & 2024 & Text & 30 & Yes & Open & \url{https://github.com/hkust-zhiyao/RTLLM/tree/v1.1}\\
    RTLLM-v2~\cite{21} & 2024 & Text & 50 & Yes & Open & \url{https://github.com/hkust-zhiyao/RTLLM/tree/main}\\
    RTLRepo\_test~\cite{102} & 2024 & Text+Code & 1.17k & No & Open & \url{https://huggingface.co/datasets/ahmedallam/RTL-Repo}\\
    ArchXBench~\cite{103} & 2025 & Text & 51 & Yes & Open & \url{https://github.com/sureshpurini/ArchXBench}\\
    CVDP~\cite{104} & 2025 & Text & 783 & Yes & Open & \url{https://github.com/NVlabs/cvdp_benchmark}\\
    GenBen~\cite{105} & 2025 & Text+Image & 324 & Yes & Open & \url{https://github.com/ChatDesignVerification/GenBen}\\
    RealBench~\cite{106} & 2025 & Text+Image & 60 & Yes & Open & \url{https://github.com/IPRC-DIP/RealBench}\\
    ResBench~\cite{107} & 2025 & Text & 56 & Yes & Open & \url{https://github.com/jultrishyyy/ResBench}\\
    VerilogEval-v2~\cite{108} & 2025 & Text & 156 & Yes & Open & \url{https://github.com/NVlabs/verilog-eval/tree/main/}\\
    VeriThoughts~\cite{87} & 2025 & Text & 291 & No & Open & \url{https://huggingface.co/datasets/wilyub/VeriThoughtsBenchmark}\\
    \hline
    \end{tabular}%
    }
\end{table*}

\begin{table*}[!t]
    \centering
    \caption{Instruct-Tuning Datasets for Verilog Code Generation (Classified by Availability). }
    \label{tab:it_datasets}
    \resizebox{\textwidth}{!}{%
    \begin{tabular}{|l|l|l|l|l|l|l|}
    \hline
    \textbf{Dataset} & \textbf{Year} & \textbf{Input} & \textbf{Samples} & \textbf{Testbench} & \textbf{Available} & \text{URL}\\
    \hline
    \multicolumn{7}{|l|}{\textbf{Closed-Source} } \\
    \hline
    DAVE\_train~\cite{9} & 2020 & Text & 5k & No & Closed & - \\
    VerilogEval\_train~\cite{57} & 2023 & Text & 8.5k & No & Closed & - \\
    BetterV~\cite{56} & 2024 & Text & Unknown & No & Closed & - \\
    MEV-LLM~\cite{63} & 2024 & Text & 31k & No & Closed & - \\
    CodeV~\cite{71} & 2025 & Text & 165k & No & Closed & - \\
    CraftRTL~\cite{58} & 2025 & Text & 80K & No & Closed & - \\
    DeepRTL~\cite{59} & 2025 & Text & 556k & No & Closed & - \\
    DeepRTL2~\cite{60} & 2025 & Text & 433k & No & Closed & - \\
    GEMMV\_train~\cite{73} & 2025 & Text & 12k & No & Closed & - \\
    OpenRTLSet~\cite{64} & 2025 & Text & 98K & No & Closed & - \\
    ScaleRTL~\cite{68} & 2025 & Text & 62k & No & Closed & - \\
    VerilogDB~\cite{109} & 2025 & Text & 20k & No & Closed & - \\
    \hline
    \multicolumn{7}{|l|}{\textbf{Open-Source} } \\
    \hline
    VeriGen\_train~\cite{88} & 2023 & Text & 109k & No & Open & \url{https://huggingface.co/datasets/shailja/Verilog_GitHub}\\
    AutoVCoder-Data~\cite{55} & 2024 & Text & 1M & No & Open & \url{https://github.com/sjtu-zhao-lab/AutoVCoder/tree/main/data}\\
    ChipGPT-FT-Data~\cite{110} & 2024 & Text & 124k & No & Open & \url{https://modelscope.cn/datasets/changkaiyan/chipgptseries}\\
    MG-Verilog-Data~\cite{76} & 2024 & Text & 11k & No & Open & \url{https://huggingface.co/datasets/GaTech-EIC/MG-Verilog}\\
    Origen-Data~\cite{78} & 2024 & Text & 222k & No & Open & \url{https://huggingface.co/datasets/henryen/origen_dataset_instruction}\\
    RTLCoder-Data~\cite{111,80} & 2024 & Text & 27K & No & Open & \url{https://github.com/hkust-zhiyao/RTL-Coder/tree/main/dataset}\\
    RTLRepo\_train~\cite{102} & 2024 & Text+Code & 2.92k & No & Open & \url{https://huggingface.co/datasets/ahmedallam/RTL-Repo}\\
    Mistral-Verilog~\cite{77} & 2024 & Text & 68k & No & Open & \url{https://huggingface.co/datasets/emilgoh/verilog-dataset-v2}\\
    CodeV-R1-Data~\cite{72} & 2025 & Text & 3k & No & Open & \url{https://huggingface.co/datasets/zhuyaoyu/CodeV-R1-dataset}\\
    DeepCircuitX~\cite{112} & 2025 & Text & 28k & No & Open & \url{https://drive.google.com/file/d/1Y002eJQPMbrEX7IpmzXDpFRlGl0XgUFu}\\
    Haven-KL~\cite{75} & 2025 & Text & 62k & No & Open & \url{https://huggingface.co/datasets/yangyiyao/HaVen-KL-Dataset}\\
    OpenCores~\cite{86} & 2025 & Text & 834 & No & Open & \url{https://huggingface.co/datasets/LLM-EDA/opencores} \\
    PyraNet~\cite{65} & 2025 & Text & 692k & No & Open & \url{https://huggingface.co/datasets/bnadimi/PyraNet-Verilog}\\
    ReasoningV-Data~\cite{79} & 2025 & Text & 5k & No & Open & \url{https://huggingface.co/datasets/GipAI/ReaoningV} \\
    RTL++-Data~\cite{66} & 2025 & Text+Graph & 11.7k & No & Open & \url{https://huggingface.co/datasets/makyash/RTL-PP}\\
    VeriCoder-Origen~\cite{81} & 2025 & Text & 126k & Yes & Open & \url{https://huggingface.co/datasets/LLM4Code/expanded_origen_126k}\\
    VeriCoder-RTLCoder~\cite{81} & 2025 & Text & 12k & Yes & Open & \url{https://huggingface.co/datasets/LLM4Code/expanded_rtlcoder_12k}\\
    VeriLogos-Data~\cite{83} & 2025 & Text & 10k & No & Open & \url{https://huggingface.co/datasets/97kjmin/VeriLogos_Augmented_Dataset}\\
    VeriPrefer-Data~\cite{84} & 2025 & Text & 90k & No & Open & \url{https://huggingface.co/datasets/LLM-EDA/pyra}\\
    VeriReason-Data~\cite{85} & 2025 & Text & 1k & Yes & Open & \url{https://huggingface.co/Nellyw888/datasets}\\
    VeriThoughts-Data~\cite{87} & 2025 & Text & 20k & No & Open & \url{https://huggingface.co/datasets/wilyub/VeriThoughtsTrainSet}\\
    \hline
    \end{tabular}%
    }
\end{table*}

\subsubsection{Benchmark Dataset Construction}\label{RQ2.1.1}

\paragraph{(1) Data Sources}
Benchmark datasets for Verilog code generation can be systematically categorized into four primary data sources according to their construction methodologies: template synthesis, mining software repositories, expert curation, and hybrid methods.

\textbf{\textcircled{1} Template Synthesis.}
Template synthesis employs structured templates with parameterized natural-language descriptions and corresponding Verilog code.
For example, \textit{DAVE}~\cite{9} constructs paired English–Verilog templates (e.g., mapping English ``OR'' to Verilog ``|'') and instantiates them via systematic parameter conversion.
This process offers strong quality control but inevitably lacks the natural diversity of real-world implementations~\cite{113}.




\textbf{\textcircled{2} Mining Software Repositories.}
This approach extracts problems and code from educational platforms and public repositories, leveraging verified solutions and broad coverage. 
\textit{VerilogEval}~\cite{57} is fully derived from HDLBits~\cite{114}: \textit{VerilogEval-machine} uses GPT-3.5-turbo to synthesize 143 problems, while \textit{VerilogEval-human} provides 156 expert-crafted problems.
\textit{VerilogEval-v2}~\cite{108} extends this with hybrid collections (HDLBits + manual curation), targeting code completion and specification-to-RTL tasks with optimized formats and in-context examples.
\textit{AutoChip}~\cite{96} also refines HDLBits into 120 executable code-generation tasks from 178 original problems and reconstructs testbenches by reverse-engineering HDLBits’ testing logic.
In contrast, \textit{RTL-repo-test}~\cite{102} mines GitHub repositories created after October 2023 with 4–24 Verilog files under permissive licenses, systematically sampling files and target lines for prediction tasks.




\textbf{\textcircled{3} Expert Curated.}
Expert-curated datasets are authored by domain specialists based on canonical designs, teaching materials, and industrial practices, typically with precise prompts and comprehensive testbenches.
Educationally oriented benchmarks such as \textit{VGV}~\cite{92} and \textit{VeriGen\_test}~\cite{88} draw on course materials and HDLBits-style exercises, covering combinational logic, sequential circuits, and finite state machines, with expert-authored diagrams and testbenches (VGV further includes visual prompts).
Standardized, broad-coverage suites like \textit{RTLLM-v1/v2}~\cite{101,21} and \textit{ChipGPTV}~\cite{97} define taxonomies across arithmetic, logic/control, storage, and advanced modules.
More complex hierarchical and application-level challenges are captured by \textit{ModelEval}~\cite{95}, \textit{CreativEval}~\cite{98}, \textit{Hierarchical}~\cite{100}, and \textit{ArchXBench}~\cite{103}, which emphasize top–submodule structures, dependency graphs, timing constraints, and diverse domains.
\textit{PyHDL-Eval}~\cite{154} further advances taxonomy-driven benchmarking with an ontological organization of 168 specification-to-RTL problems across 19 RTL design subcategories, each shipped with Verilog reference solutions and testbenches, enabling finer-grained diagnosis of LLM capabilities across design categories.
Application-oriented benchmarks such as \textit{GEMMV}~\cite{73} and \textit{CVDP}~\cite{104} focus on AI accelerators and design/verification scenarios, respectively, while \textit{ResBench}~\cite{107} targets real FPGA applications (e.g., ML acceleration, finance, encryption) with an emphasis on resource optimization.
\textit{HiVeGen}~\cite{94} further supports accelerator evaluation via LLM-based kernel analysis and dataflow graph extraction.

\textbf{\textcircled{4} Hybrid Methods.}
Hybrid methods combine open-source mining with expert refinement or LLM-based synthesis to improve both coverage and data quality.
\textit{AutoSilicon}~\cite{93} mines open-source repositories for industrial modules (I/O controllers, processors, accelerators) and augments them with expert-written prompts aligned with RTLLM description formats.
\textit{GenBen}~\cite{105} integrates verification projects, textbooks, Stack Overflow discussions, and open-source hardware resources, organizing them into knowledge, design, debugging, and multimodal categories, and applies data perturbation to avoid contamination and improve robustness.
\textit{RealBench}~\cite{106} builds on four verified open-source IPs (e.g., AES cores, SD controllers, Hummingbirdv2 E203 CPU), providing expert-optimized specifications, 100\% line-coverage testbenches, and formal verification for both module- and system-level tasks.
\textit{VeriThoughts}~\cite{87} combines the MetRex dataset~\cite{115} (25.8K synthesizable Verilog designs) with LLM-generated prompts, reasoning traces, and candidate Verilog code (using models such as Gemini and DeepSeek-R1), followed by formal checking and human sampling to ensure correctness.

\paragraph{(2) Quality Assurance}
Quality control for benchmark datasets centers on three complementary approaches that ensure reliability and correctness.

\textbf{\textcircled{1} Expert Curation and Standardization.}
Expert curation and standardization reduce ambiguity and enhance reproducibility. 
Benchmarks adopt expert-authored specifications and reviews, along with unified artifacts that enforce consistent interfaces and evaluation through standardized prompt/module-header formats or structured, annotated diagrams~\cite{101,21,96,90,97,106,104,92,88}.


\textbf{\textcircled{2} Testbench and Formal/Toolchain Checks.}
Testbench and formal/toolchain checks verify functional correctness and implementation viability. 
Existing work ranges from reverse-engineered testbenches where platform logic is undisclosed~\cite{96} to coverage-driven or 100\% line-coverage suites~\cite{105,106}, hierarchical unit–integration testing~\cite{100}, and multi-scenario/boundary testing with explicit timing constraints~\cite{95}. 
Formal methods complement simulation via equivalence checking and SAT-based proofs with self-consistency verification~\cite{106,87}, while EDA pipelines (e.g., Verilator, Yosys, Design Compiler) provide linting, synthesis, and PPA analysis~\cite{70,94}.


\textbf{\textcircled{3} Contamination Control.}
Contamination control preserves evaluation fairness. 
Typical strategies include decontamination against external corpora, static/dynamic perturbations with expert verification, multi-stage quality checks assisted by LLM-judge filters with detailed logs to guard against data leakage and noise~\cite{106,105,104,21}.


\subsubsection{Instruct-Tuning Dataset Construction}\label{RQ2.1.2}

\paragraph{(1) Data Sources}
Instruct-tuning datasets for Verilog code generation can be systematically categorized into three primary data-source strategies: mining software repositories, LLM synthesis, and hybrid methods.

\textbf{\textcircled{1} Mining Software Repositories.}
Similar to benchmark datasets, this strategy exclusively collects data from public repositories, educational platforms, and community resources without relying on LLM-generated augmentation. 
\textit{VeriGen\_train}~\cite{88} integrates GitHub mining with educational resources, processing 70 Verilog textbooks via OCR and text extraction to build a 400MB training corpus alongside GitHub-derived modules. 
\textit{VerilogDB}~\cite{109} and \textit{RTL-repo}~\cite{102} extend this approach by incorporating GitHub, OpenCores~\cite{116}, and academic sources, organizing data into structured databases with rich metadata. 
This methodology guarantees authenticity and legal compliance but requires heavy preprocessing to handle quality variation and lacks the semantic consistency of LLM-generated descriptions.




\textbf{\textcircled{2} LLM Synthesis.}
LLM synthesis generates training data entirely via language models using instruction-following and structured synthetic data generation. 
\textit{RTLCoder}~\cite{111,80} exemplifies this with a curated pool of 350 hardware design keywords across more than 10 circuit categories, using GPT-3.5 to create over 27,000 instruction–code pairs. 
\textit{CraftRTL}~\cite{58} adopts Self-Instruct, OSS-Instruct, and Docu-Instruct pipelines to synthesize 80.1k samples, addressing non-textual representations through careful prompt design. 
\textit{ScaleRTL}~\cite{68} leverages DeepSeek-R1 to produce 3.5B-token Chain-of-Thought datasets via automated specification generation and multi-step reasoning synthesis. 
This strategy offers excellent scalability and consistency, but may under-represent the diversity and practical quirks of human-authored RTL.




\textbf{\textcircled{3} Hybrid Methods.}
Hybrid methods start from community-sourced code and apply LLM-based enhancement for description generation, quality improvement, or augmentation. 
\textit{VerilogEval-train}~\cite{57} mines GitHub Verilog modules and uses GPT-3.5-turbo to generate 8,502 instruction–code pairs with systematic validation; \textit{BetterV}~\cite{56} combines authentic GitHub data with LLM-synthesized “virtual” data, using V2C translation tools and EDA verification to guarantee functional correctness; \textit{MEV-LLM}~\cite{63} merges GitHub mining with ChatGPT-3.5-Turbo descriptions to create 31,104 samples labeled across four complexity levels. 
Multi-source integration is further illustrated by \textit{AutoVCoder}~\cite{55}, which collects 100,000 RTL modules from GitHub and employs ChatGPT-3.5 for problem–code pair generation with filtering and verification via Icarus Verilog and Python equivalence tests; \textit{MG-Verilog}~\cite{76} augments mined repositories with multi-granularity descriptions (from line-level comments to full summaries) using LLaMA2-70B-Chat and GPT-3.5; and \textit{Origen}~\cite{78} performs code-to-code augmentation on open-source RTL, with Claude3-Haiku generating descriptions and regenerated code under iterative compilation checks. 
Expert knowledge integration appears in \textit{Haven}~\cite{75}, which combines GitHub code with textbook examples and professional annotations, and in \textit{DeepRTL}~\cite{59}, which fuses GitHub and proprietary industrial IPs with GPT-4 Chain-of-Thought annotation and professional verification, achieving 90\% annotation precision under human evaluation.

\paragraph{(2) Quality Assurance}
Quality control for instruct-tuning datasets converges on five complementary approaches that ensure training data reliability and optimize model performance.

\textbf{\textcircled{1} Syntax and Synthesizability Verification.}
This step checks that training samples satisfy basic hardware design requirements via automated toolchains. 
\textit{VeriGen\_train}~\cite{88} uses Pyverilog for AST extraction and syntax validation, filtering modules to guarantee complete \texttt{module}–\texttt{endmodule} structures and proper port definitions. 
\textit{AutoVCoder}~\cite{55} performs Icarus Verilog compilation followed by Yosys synthesis to ensure mappability to gate-level netlists, removing non-synthesizable constructs (e.g., \texttt{initial} blocks, \texttt{\$display}). 
\textit{VeriCoder}~\cite{81} combines syntax validation with executable testbenches using iverilog to ensure both syntactic and basic behavioral correctness, while \textit{DeepRTL}~\cite{59} validates tokenizer compatibility (CodeT5+, 2048-token context) and structural integrity via Pyverilog.

\textbf{\textcircled{2} Functional Correctness Verification.}
Functional verification extends beyond syntax to confirm that code produces expected outputs. 
\textit{VeriCoder}~\cite{81} uses GPT-4o to generate unit-test style testbenches (input stimuli and assertions) and verifies equivalence between generated and reference implementations with iverilog. 
\textit{VeriThoughts}~\cite{87} conducts formal equivalence checking with Yosys \texttt{miter} and SAT solvers, guaranteeing identical outputs across all inputs. 
\textit{ReasoningV}~\cite{79} augments simulation with boundary-case analysis (edge cases, reset, error conditions), and \textit{VeriPrefer}~\cite{84} applies coverage-driven testing via VCS, achieving over 90\% line coverage and assertion validation.

\textbf{\textcircled{3} Contamination Control.}
Contamination control preserves dataset integrity by removing redundancy and preventing leakage from evaluation benchmarks. 
\textit{PyraNet}~\cite{65} uses Jaccard similarity on tokenized code, discarding samples above 0.9 similarity to avoid overfitting to near-duplicates. 
\textit{CodeV}~\cite{71} adopts MinHash-based deduplication with 128-dimensional vectors to filter duplicates while maintaining diversity. 
\textit{VeriLogos}~\cite{83} compares training data against public benchmarks (VerilogEval, RTLLM) using Rouge-L, removing samples with similarity over 0.5, and \textit{ScaleRTL}~\cite{68} uses 5-gram sequence analysis against benchmark solutions to detect and eliminate contamination.

\textbf{\textcircled{4} Format Standardization and Length Control.}
Standardization ensures consistent representation and compatibility with model constraints. 
\textit{Haven}~\cite{75} normalizes naming conventions (e.g., \texttt{clk}, \texttt{rst\_n}), module declarations, and indentation (4-space), while pruning redundant comments and keeping essential documentation. 
\textit{OpenRTLSet}~\cite{64} uses the CodeLlama tokenizer to enforce an 8k-token limit, and \textit{VeriReason}~\cite{85} constrains specification descriptions (100–300 words), reasoning (150–300 words), and code (<2048 tokens) for consistent input formatting. 
\textit{RTL++}~\cite{66} extends standardization to multimodal data, enforcing structured instruction–code–graph triplets and consistent CFG/DFG representations within context limits.

\textbf{\textcircled{5} Quality Filtering and Human Validation.}
Automated filtering is complemented by expert review to ensure annotation accuracy and domain correctness. 
\textit{PyraNet}~\cite{65} employs GPT-4o-mini for 0–20 quality scoring (style, efficiency, standardization), followed by hardware-engineer validation. 
\textit{DeepRTL}~\cite{59} conducts multi-stage expert review with professional Verilog designers and additional engineers, achieving 90\% annotation accuracy. 
\textit{VeriCoder}~\cite{81} combines automatic checks with engineer validation of specification–code consistency and testbench effectiveness, reporting 92\% functional correctness, while \textit{Haven}~\cite{75} leverages textbook examples and professional annotations to build knowledge- and logic-enhanced datasets aligned with engineering practice.

These mechanisms collectively provide reliable, diverse, and functionally correct supervision for instruct-tuning, while mitigating data leakage, annotation errors, and functional inconsistencies.

\subsubsection{Trends Analysis}
We highlight three key evolutionary aspects of datasets and their implications for evaluation: (1) testbench availability as a proxy for executability, (2) input modality as benchmarks move beyond text-only specifications, and (3) dataset scale as tasks range from reasoning-focused micro-suites to broad coverage collections.

\paragraph{(1) Testbench availability} 
Executable verification has effectively become the standard for evaluation. 
Among open benchmarks, 17 of 19 provide testbenches (exceptions: \textit{RTLRepo\_test}, which targets line-level generation, and \textit{VeriThoughts}, which uses Yosys-based formal equivalence instead of simulation). 
In the closed-source category, all datasets except \textit{DAVE\_test} include testbenches. 
By contrast, most instruct-tuning datasets lack executable supervision; only a few (e.g., \textit{VeriCoder-Origen}, \textit{VeriCoder-RTLCoder}, \textit{VeriReason-Data}) incorporate testbenches or similar signals. 
This gap underscores the need to introduce executable supervision (testbenches, formal specifications, synthesis feedback) earlier in training pipelines.

\paragraph{(2) Input modality} 
While input specifications remain predominantly text-based, benchmarks are increasingly multimodal (Text+Image), as seen in \textit{ChipGPTV}~\cite{97}, \textit{VGV}~\cite{92}, \textit{GenBen}~\cite{105}, and \textit{RealBench}~\cite{106}, reflecting realistic workflows where timing diagrams, schematics, and waveforms complement textual requirements. 
In training resources, instruction corpora are still mostly text-only, with emerging structural representations (e.g., Text+Graph in \textit{RTL++-Data}) that expose hierarchy and connectivity crucial for RTL reasoning.

\paragraph{(3) Dataset scale} 
Dataset scale diverges with purpose. 
Benchmarks range from compact reasoning-oriented suites (4–30 items) to mid/large-scale resources (e.g., \textit{CVDP}: 783; \textit{GenBen}: 324) emphasizing coverage and diversity. 
Instruction-tuning datasets span from specialized compact sets (e.g., \textit{CodeV-R1-Data}) to hundred-thousand or million-scale corpora (e.g., \textit{AutoVCoder-Data}, \textit{PyraNet}), where the former sharpen reasoning and alignment and the latter support broad generalization despite weaker executable validation. 
This complementarity between small, high-signal benchmarks and large, weakly validated training sets mirrors broader practice in code-oriented LLMs.

Based on these observations, future dataset development should prioritize: (1) early integration of \textit{executable supervision} to narrow the gap between training and evaluation; (2) stronger \textit{multimodal alignment} between training and benchmark specifications; and (3) \textit{reasoning-enhanced training} that explicitly encodes intermediate reasoning and design decisions, as explored by CodeV-R1~\cite{72}, VeriReason~\cite{85}, ScaleRTL~\cite{68}, and VeriThoughts~\cite{87}.

\subsection{Evaluation Metrics}\label{RQ2.2}
Assessing the quality of LLM-generated Verilog code is challenging, as it must account for both syntactic validity and semantic correctness. 
Existing work can be broadly grouped into three categories: similarity-based, execution-based, and LLM-based metrics.

\subsubsection{Similarity-based Metrics}\label{RQ2.2.1}
Similarity-based metrics evaluate generated Verilog by comparing it to reference implementations without executing the code.

\paragraph{(1) General-Purpose Metrics} 
Most studies adopt text/code similarity metrics from NLP and general code evaluation:
\begin{itemize}
\item \textbf{Exact Match:} Binary check of whether generated and reference code are identical~\cite{102,67}
\item \textbf{Edit Distance Metrics:} Character-level alignment via Levenshtein distance~\cite{102,67} and Ratcliff-Obershelp similarity~\cite{9}
\item \textbf{N-gram Based Metrics:} Sequence-overlap measures such as BLEU~\cite{60,104,112,117}, METEOR~\cite{60,112}, ROUGE~\cite{60,112,117}, and chrF~\cite{117}, originating from machine translation
\end{itemize}
These metrics are computationally efficient and easy to implement, but only approximate functional correctness.

\paragraph{(2) Verilog-Specific Metrics} 
To better capture hardware-specific semantics, SimEval~\cite{118} introduces a Verilog-oriented metric operating on three levels: (1) \textit{Syntactic} analysis via PyVerilog ASTs; (2) \textit{Semantic} analysis using CFGs extracted by Verilator; and (3) \textit{Functional} analysis through gate-level netlist comparison after Yosys synthesis. 
This multi-view design targets RTL-specific characteristics such as hierarchy, concurrency, and register-transfer behavior.

\subsubsection{Execution-based Metrics}\label{RQ2.2.2}
Execution-based metrics evaluate generated code through compilation and runtime checks, directly measuring functional correctness and implementation quality.

\paragraph{(1) Correctness Assessment} 
Core execution-based metrics include:
\begin{itemize}
\item \textbf{Syntax-pass@k:} Fraction of samples that compile without syntax errors within \(k\) attempts
\item \textbf{Functional-pass@k:} Fraction of samples that pass standardized testbenches with expected outputs
\item \textbf{Formal Verification:} Equivalence checking between generated and reference code using tools such as \texttt{Yosys -equiv}
\end{itemize}

\paragraph{(2) Hardware Quality Metrics} 
Beyond correctness, some work evaluates hardware implementation quality via PPA (Power, Performance, Area) analysis:
\begin{itemize}
\item \textbf{Power:} Energy efficiency of synthesized designs
\item \textbf{Performance:} Timing metrics such as maximum frequency, critical path delay, and latency
\item \textbf{Area:} Resource utilization in terms of logic, storage elements, and overall silicon footprint
\end{itemize}

\subsubsection{LLM-based Metrics}\label{RQ2.2.3}
LLM-based metrics leverage language models themselves to capture deeper semantics and domain knowledge.

\paragraph{(1) Model Confidence Metrics} 
These metrics analyze internal confidence during generation:
\begin{itemize}
\item \textbf{Perplexity:} Measures model certainty over generated Verilog sequences~\cite{117,74}
\end{itemize}

\paragraph{(2) LLM-as-a-Judge} 
Here, LLMs act as expert evaluators on multiple quality dimensions:
\begin{itemize}
\item \textbf{GPT Score:} Uses GPT-family models to rate code on readability, correctness, and design best practices~\cite{59,60}
\item \textbf{VCD-RNK:} Assesses semantic consistency between natural-language specifications and implementations, with knowledge distillation to obtain lightweight evaluators~\cite{119}
\item \textbf{MetRex:} Employs LLMs as PPA predictors for Verilog designs~\cite{115}
\end{itemize}

\begin{tcolorbox}[left=4pt,right=4pt,top=2pt,bottom=2pt,boxrule=0.5pt]
  \textbf{Answer to RQ2:} 
  (1) Verilog datasets fall into two main categories: benchmarks and instruct-tuning corpora. They are constructed via template synthesis, open-source mining, expert curation, LLM-based synthesis, and hybrid pipelines, with quality safeguarded by executable testbenches, formal verification, contamination control, and standardized formats. 
  (2) Evaluation metrics have evolved from text similarity measures toward execution-based assessment, with functional-pass@k becoming the dominant indicator of correctness. Future work should further integrate hardware quality metrics (PPA) and develop LLM-as-a-Judge frameworks for reference-free, semantically faithful evaluation.
\end{tcolorbox}

%% file: sec/06_RQ3.tex
\section{RQ3: Adaptation and Optimization Techniques for Verilog Code Generation}\label{RQ3}
Building on the datasets and evaluation methodologies discussed in the previous sections, we now examine how LLMs are adapted and optimized for Verilog code generation. 
These techniques can be broadly divided into training-free and training-based approaches, each with distinct trade-offs between computational efficiency and performance gains.

\subsection{Training-free Methods}\label{RQ3.1}
Training-free methods enhance Verilog generation without additional parameter updates, making them especially attractive when computational resources are limited. 
They mainly rely on EDA-tool feedback, prompt engineering, and inference-time decoding optimization.

\subsubsection{EDA-Tool Feedback}\label{RQ3.1.1}
Industry-standard verification and synthesis tools provide diagnostics (syntax, functional, timing, PPA) that close the loop from generation to refinement. Approaches split into single-agent (Table~\ref{tab:single_agent}) and multi-agent (Table~\ref{tab:multi_agent}) frameworks.

\paragraph{(1) Single-Agent Systems}\label{RQ3.1.1.1}
Table~\ref{tab:single_agent} contrasts six representative systems. They share a staged feedback loop (correctness first, PPA second) but differ in \emph{how} feedback is consumed: role prompts, ReAct repair, paradigm decomposition, evolutionary search, or multimodal input.

\begin{table}[!t]
\centering
\caption{Comparison of single-agent EDA-feedback systems.}
\label{tab:single_agent}
\resizebox{\columnwidth}{!}{
\begin{tabular}{llll}
\toprule
\textbf{System} & \textbf{Feedback Source} & \textbf{Refinement Strategy} & \textbf{Optimization Target} \\
\midrule
VeriPPA~\cite{133} & iverilog + Synopsys DC & Dual-stage pipeline & Correctness + PPA \\
RTL Assistant~\cite{138} & Icarus Verilog & Role-based prompts (3 types) & Correctness \\
RTLFixer~\cite{139} & Compiler + RAG & ReAct think-act-observe loops & Syntax repair \\
Paradigm-Based~\cite{137} & PyEDA + Simulator & Paradigm block decomposition & Correctness \\
EvoVerilog~\cite{140} & Testbench + Synthesis & Evolutionary search + non-dominated sorting & Multi-objective \\
VGV~\cite{92} & Simulator & Multimodal visual prompting & Correctness \\
\bottomrule
\end{tabular}}
\end{table}

\begin{table*}[!t]
\centering
\caption{Comparison of multi-agent EDA-feedback systems for Verilog generation.}
\label{tab:multi_agent}
\resizebox{\textwidth}{!}{
\begin{tabular}{lclll}
\toprule
\textbf{System} & \textbf{\#Agents} & \textbf{Architecture Type} & \textbf{EDA Tools} & \textbf{Primary Focus} \\
\midrule
AIVRIL~\cite{142} & 2 & Generator-Reviewer & Compiler logs & Syntax + functional correctness \\
LLM-aided Front-End~\cite{89} & 3 & Sequential pipeline & Behavioral simulator & RTL + testbench + review \\
RTL Agent~\cite{143} & 2 & Generator-Reviewer (Reflexion) & Testbench & Parallel exploration vs. depth \\
\midrule
MAGE~\cite{144} & 4 & Role-specialized team & Simulator & Scoring + iterative debug \\
VerilogCoder~\cite{145} & 3+ & Task planner + verifier + debugger & iverilog + Pyverilog AST & Waveform-level mismatch localization \\
CoopetitiveV~\cite{146} & 4 & Competitive-cooperative & iverilog + AutoBench & Adversarial refinement \\
\midrule
Nexus~\cite{147} & Hierarchical & Layered supervisor & Xilinx Vivado & Timing + resource + power \\
RTLSquad~\cite{134} & 3 squads & Squad-based (explore/implement/verify) & EDA reports & Interpretable PPA optimization \\
VFlow~\cite{148} & Population & Evolutionary MCTS & Multiple operators & Automated workflow discovery \\
\midrule
VeriOpt~\cite{131} & 4 & Planner-Programmer-Reviewer-Evaluator & Synopsys DC & PPA-aware two-stage iteration \\
AutoSilicon~\cite{93} & 3 & Designer + Debugger + Scheduler & iverilog + Synopsys DC & Industrial-scale RTL \\
VeriMind~\cite{149} & 5 & Full-pipeline collaboration & EDA integration & pass@ARC metric \\
\bottomrule
\end{tabular}}
\end{table*}

Overall, single-agent systems demonstrate that carefully engineered feedback loops, whether staged, iterative, or evolutionary, can substantially improve correctness and PPA without requiring additional model instances.

\paragraph{(2) Multi-Agent Systems}\label{RQ3.1.1.2}
Table~\ref{tab:multi_agent} reveals a consistent architectural progression: Generator--Reviewer pairs~\cite{142,143} $\rightarrow$ role-specialized teams with competitive verification~\cite{144,145,146} $\rightarrow$ hierarchical supervisors and automated workflow search~\cite{147,148}, culminating in industrial PPA-aware pipelines~\cite{131,93,149}. The recurring design principle is separation of concerns: generation, verification, and optimization are handled by distinct agents rather than a single monolithic loop.

\subsubsection{Prompt Engineering}\label{RQ3.1.2}
Two training-free strategies stand out. \textit{Structured prompting} decomposes specifications into intermediate representations before Verilog generation~\cite{95,94,120}. \textit{RAG} injects domain knowledge at inference time~\cite{121,122}; VeriGRAG~\cite{155} extends this with GNN-derived structural embeddings aligned to the LLM via soft prompts.

\subsubsection{Inference Optimization}\label{RQ3.1.3}
Decoding-time reranking improves candidate quality without parameter updates: syntax-aware temperature scaling~\cite{123}, execution-based self-consistency clustering~\cite{124}, and distilled semantic reranking~\cite{119}.

\subsection{Training-based Methods}\label{RQ3.2}
Training-based methods adapt model parameters and generally outperform training-free approaches at higher compute cost.

\subsubsection{Pre-training}\label{RQ3.2.1}
Two strategies exist: training from scratch (e.g., Hardware Phi-1.5B~\cite{74}) and continued pre-training of general LLMs on Verilog corpora (e.g., FreeV~\cite{61} via LoRA).

\subsubsection{Supervised Fine-tuning}\label{RQ3.2.2}
SFT minimizes $\mathcal{L}_{SFT} = -\sum_{(x,y)} \sum_i \log P(y_i | x, y_{<i}; \theta)$ on specification--code pairs using full fine-tuning, LoRA, or QLoRA. Table~\ref{tab:sft_summary} lists one row per system; the discussion below highlights cross-cutting patterns only.

\begin{table*}[!t]
\centering
\caption{Comparison of supervised fine-tuning approaches for Verilog generation (one row per system).}
\label{tab:sft_summary}
\resizebox{\textwidth}{!}{
\begin{tabular}{lllll}
\toprule
\textbf{System} & \textbf{Year} & \textbf{Tuning} & \textbf{Key Technique} & \textbf{Dataset (size)} \\
\midrule
\multicolumn{5}{l}{\textit{Data-Centric --- compete on dataset construction}} \\
DAVE~\cite{9}          & 2020 & Full SFT       & English--Verilog template synthesis              & DAVE (5k) \\
VeriGen~\cite{88}      & 2023 & Full SFT       & GitHub + textbook mining                         & VeriGen (109k) \\
RTLCoder~\cite{80}     & 2024 & Multi-task LoRA& Keyword-driven LLM synthesis                     & RTLCoder (27k) \\
AutoVCoder~\cite{55}   & 2024 & LoRA           & Hybrid GitHub + LLM pairs                        & AutoVCoder (1M) \\
MG-Verilog~\cite{76}   & 2024 & QLoRA          & Multi-granularity descriptions                   & MG-Verilog (11k) \\
CraftRTL~\cite{58}     & 2025 & ---            & Self-/OSS-/Docu-Instruct synthesis               & CraftRTL (80k) \\
\midrule
\multicolumn{5}{l}{\textit{Strategy-Centric --- reshape how a fixed corpus is consumed}} \\
MEV-LLM~\cite{63}      & 2024 & Full SFT       & Complexity-routed expert ensemble                & MEV-LLM (31k) \\
DeepRTL~\cite{59}      & 2025 & Full SFT       & 3D curriculum (structure/granularity/quality)   & DeepRTL (556k) \\
PyraNet~\cite{65}      & 2025 & LoRA           & Hierarchical loss weighting (6 quality layers)  & PyraNet (692k) \\
\midrule
\multicolumn{5}{l}{\textit{Multi-task --- add auxiliary supervision to generation}} \\
BetterV~\cite{56}      & 2024 & Multi-task LoRA& Generator--Discriminator                         & BetterV \\
OriGen~\cite{78}   & 2024 & Multi-task LoRA& Gen-LoRA + Fix-LoRA (gen--repair)               & OriGen (222k) \\
DeepRTL2~\cite{60}     & 2025 & Full SFT       & Generation + contrastive loss                    & DeepRTL2 (433k) \\
ITERTL~\cite{62}       & 2025 & ---            & Generation + ranking loss                        & ITERTL \\
CodeV~\cite{71}        & 2025 & Full SFT       & Generation + Fill-in-Middle                      & CodeV (165k) \\
\midrule
\multicolumn{5}{l}{\textit{Knowledge-Enhanced --- inject structure or reasoning}} \\
RTL++~\cite{66}        & 2025 & LoRA           & CFG/DFG structural injection                     & RTL++ (11.7k) \\
VeriThoughts~\cite{87} & 2025 & Full SFT       & DeepSeek-R1 reasoning distillation               & VeriThoughts (20k) \\
CodeV-R1~\cite{72}     & 2025 & DAPO           & Reasoning distillation + DAPO                    & CodeV-R1 (3k) \\
ScaleRTL~\cite{68}     & 2025 & ---            & CoT reasoning synthesis                          & ScaleRTL (62k) \\
ReasoningV~\cite{79}   & 2025 & LoRA + Full SFT& Adaptive reasoning depth                         & ReasoningV (5k) \\
\bottomrule
\end{tabular}}
\end{table*}

\paragraph{(1) Data-Centric Tuning}\label{RQ3.2.2.1}
Innovation lies in \emph{what} is trained on: single-source corpora give way to hybrid pipelines with quality control. On well-curated data, LoRA matches full SFT at lower cost---data quality, not the update strategy, is the primary lever.

\paragraph{(2) Strategy-Centric Tuning}\label{RQ3.2.2.2}
Innovation lies in \emph{how} a fixed corpus is consumed: complexity routing, curriculum ordering, or hierarchical loss weighting each extract gains that plain SFT misses.

\paragraph{(3) Multi-task Tuning}\label{RQ3.2.2.3}
Multi-task tuning (the Multi-task block of Table~\ref{tab:sft_summary}) adds auxiliary supervision to the generation objective along two axes. \textit{Architecturally}, dual-task frameworks pair generation with a complementary capability---quality discrimination (BetterV~\cite{56}) or compiler-feedback-driven repair (OriGen~\cite{78,110}). 
\textit{In the loss space}, auxiliary objectives such as contrastive loss (RTLCoder~\cite{80}, DeepRTL2~\cite{60}), ranking loss (ITERTL~\cite{62}), decoding constraints~\cite{125}, or Fill-in-Middle~\cite{71} are added to the cross-entropy term. 
The common motivation is mitigating exposure bias and broadening downstream coverage; the evidence suggests these signals are most effective when they encode task structure (e.g., repair or ranking) that plain next-token prediction cannot capture.

\paragraph{(4) Knowledge-enhanced Tuning}\label{RQ3.2.2.4}
External knowledge enters via CFG/DFG injection or, more recently, reasoning-trace distillation. The latter dominates on complex designs and directly enables the self-correction behaviors exploited by RL (Section~\ref{RQ3.2.3}).

Overall, the field progresses from competing on data (2020--2024) to competing on training strategy and auxiliary supervision, with the 2025 cohort converging on reasoning distillation plus parameter-efficient tuning---the strongest SFT axis to date and the natural springboard for RL.

\subsubsection{Reinforcement Learning}\label{RQ3.2.3}

While supervised approaches rely on static datasets, reinforcement learning methods enable dynamic optimization through interaction with the environment, offering potential for continuous improvement in code generation quality. 
Reinforcement learning (RL) enables dynamic optimization by training a policy $\pi_\theta$ to maximize expected rewards $R$:
\begin{equation}
    J(\theta) = \mathbb{E}_{x \sim p_{data}, y \sim \pi_\theta(\cdot|x)}[R(y, y^*)]
\end{equation}
where $x$ represents the natural language specification, $y$ denotes the generated Verilog code, $y^*$ is the reference implementation, and $R(y, y^*)$ quantifies code quality through various metrics including functional correctness and structural similarity.
Current approaches are categorized by their reward mechanisms:

\paragraph{(1) Structure-based Rewards.}\label{RQ3.2.3.1} 
To improve syntactic quality, methods like VeriSeek~\cite{86} (using PPO) and VeriReason~\cite{85} (using GRPO) incorporate Abstract Syntax Tree (AST) similarity into the reward function. 
These approaches typically employ LoRA for efficient iterative optimization, penalizing invalid syntax while rewarding structural alignment with reference code.


\paragraph{(2) Tool-based Feedback.}\label{RQ3.2.3.2} 
Leveraging EDA tools provides objective correctness signals. VeriLogos~\cite{83} utilizes formal equivalence checking, while VeriPrefer~\cite{84} and CodeV-R1~\cite{72} employ Direct Preference Optimization (DPO/DAPO) based on testbench results. 
By marking code that passes more tests as preferred ($y^+$), these methods align model outputs with functional verification tools:
\begin{equation}
    \mathcal{L}_{DPO} = -\mathbb{E}[\log \sigma(\beta \log \frac{\pi_\theta(y^+|x)}{\pi_{ref}(y^+|x)} - \beta \log \frac{\pi_\theta(y^-|x)}{\pi_{ref}(y^-|x)})]
\end{equation}


\paragraph{(3) Multi-objective Optimization.}\label{RQ3.2.3.3} Beyond correctness, frameworks like MCTS-augmented generation~\cite{90} model RTL design as a Markov Decision Process. They integrate Power, Performance, and Area (PPA) metrics into the reward, enabling lookahead-guided decoding that optimizes for downstream hardware constraints.


Empirical evidence demonstrates that RL-based strategies consistently outperform supervised baselines by directly optimizing for the final design goals, whether syntax, functional correctness, or hardware efficiency.


\begin{tcolorbox}[left=4pt,right=4pt,top=2pt,bottom=2pt,boxrule=0.5pt]
    \textbf{Answer to RQ3:} 
    (1) Training-free methods include EDA tool feedback systems (encompassing single-agent iterative refinement and multi-agent collaborative architectures), prompt engineering strategies (hierarchical prompting and RAG-enhanced approaches), and inference optimization techniques (syntax-aware decoding and self-consistency reranking). 
    (2) Training-based methods span pre-training (from-scratch and continued pre-training), supervised fine-tuning (data-centric, strategy-centric, multi-task, and knowledge-enhanced approaches), and reinforcement learning (structure-based rewards, tool-based feedback, and multi-objective optimization).     
\end{tcolorbox}

%% file: sec/07_RQ4.tex
\section{RQ4: Alignment Approaches for Verilog Code Generation}\label{RQ4}
While RQ3 highlighted the capabilities of LLMs in Verilog generation, ensuring these models align with human intentions and rigorous hardware requirements remains critical. 
Unlike software, hardware design faces strict physical constraints, safety imperatives, and commercial IP risks. Consequently, alignment in this domain encompasses four essential dimensions:

\begin{itemize}
    \item \textbf{Security:} Guaranteeing generated code is free from vulnerabilities and hardware trojans that could compromise system integrity.
    \item \textbf{Efficiency:} Optimizing for Power, Performance, and Area (PPA) to prevent resource wastage and ensure operational viability.
    \item \textbf{Copyright:} Mitigating IP infringement risks by ensuring generated designs do not violate existing patents or licenses.
    \item \textbf{Hallucinations:} Minimizing plausible but functionally incorrect outputs that manifest as logical inconsistencies during implementation.
\end{itemize}

The following sections examine current methodologies and challenges in addressing these alignment principles.

\subsection{Security}\label{RQ4.1}
Security in LLM-generated Verilog faces two main threats: unintentional vulnerabilities from limited domain knowledge, and intentional backdoors from data poisoning.


\paragraph{(1) Current Methods} 
Research addresses these via mitigation and defense strategies:
\begin{itemize}
    \item \textbf{Vulnerability Mitigation:} To prevent unintentional flaws, frameworks augment LLMs with security knowledge graphs. SecFSM~\cite{126} targets Finite State Machines by retrieving remediation strategies for specific vulnerabilities (e.g., Dead States). Similarly, SecV~\cite{127} constructs a Hardware-CWE knowledge graph to enable targeted retrieval and verification of security constraints during generation.
    \item \textbf{Threat Defense:} Addressing external threats involves detecting attacks and cleansing models. RTL-Breaker~\cite{128} exposes how training data poisoning can introduce backdoors triggered by specific patterns. Conversely, VeriContaminated~\cite{129} detects benchmark contamination in training sets to ensure evaluation integrity. For remediation, SALAD~\cite{130} applies machine unlearning to remove sensitive information or malicious patterns while preserving core model utility.
\end{itemize}

\paragraph{(2) Challenges \& Future Directions}
Current frameworks remain inadequate for sophisticated hardware threats.
\begin{itemize}
    \item \textbf{Limitations:} \textcircled{1} Existing knowledge integration (e.g., CWE) often lacks depth for circuit-level vulnerabilities; \textcircled{2} defense research lags behind attack methodologies; and \textcircled{3} risks like jailbreaking or adversarial attacks remain underexplored in the hardware domain.
    \item \textbf{Solutions:} \textcircled{1}\textbf{Holistic Evaluation:} Developing frameworks beyond CWE to cover hardware-specific patterns.
\textcircled{2}\textbf{Multi-stage Verification:} Combining static analysis, formal verification, and simulation to catch subtle flaws.
\textcircled{3} \textbf{Adversarial Hardening:} Fine-tuning with adversarial examples to improve robustness.
\textcircled{4} \textbf{Hardware Red-Teaming:} Collaborations between security and AI experts to systematically uncover vulnerabilities.
\end{itemize}

\subsection{Efficiency}\label{RQ4.2}
Optimizing Power, Performance, and Area (PPA) is crucial for hardware viability. Unlike software, physical constraints necessitate rigorous optimization strategies.


\paragraph{(1) Current Methods}
Research aligns LLM outputs with PPA goals through three main strategies:
\begin{itemize}
    \item \textbf{Design Space Exploration:} Systematic search identifies optimal implementations. Methods range from MCTS-guided generation (Delorenzo et al.~\cite{90}) that rewards PPA efficiency, to ChipGPT's~\cite{70} enumerative selection from multiple candidates.
    \item \textbf{Knowledge-Enhanced Optimization:} Injecting domain expertise improves results. VeriOpt~\cite{131} uses in-context learning with synthesis reports to select optimization techniques, while Chang et al.~\cite{132} leverage program analysis and ``Chip Discovery Search'' to combine design strengths.
    \item \textbf{Design Decomposition:} Breaking down complexity facilitates optimization. Hierarchical approaches like HiVeGen~\cite{94} and AutoSilicon~\cite{93} target submodules, while sequential frameworks (LLM-VeriPPA~\cite{133}, RTLSquad~\cite{134}) separate functional verification from PPA refinement or use specialized agents for distinct objectives.
\end{itemize}

\paragraph{(2) Challenges \& Future Directions}
Despite these advances, several challenges persist in aligning LLM-generated Verilog code with efficiency requirements. 

\begin{itemize}
    \item \textbf{Limitations:} 
    \textcircled{1} LLMs inherently lack physical design awareness: they cannot reliably predict how RTL transformations affect post-synthesis PPA metrics without external EDA feedback;
    \textcircled{2} even given such metrics, they struggle to reason about competing PPA trade-offs (e.g., pipelining improves performance but increases area);
    and \textcircled{3} Context-dependent constraints and the high computational cost of search-based methods further complicate alignment.
    \item \textbf{Solutions:} 
    \textcircled{1}\textbf{Multi-Objective Modeling:} Explicitly modeling trade-offs to balance competing constraints.
    \textcircled{2}\textbf{Closed-Loop Feedback:} Integrating EDA tools directly into generation loops to provide concrete PPA signals.
    \textcircled{3}\textbf{Specialized Datasets:} Curating data that pairs descriptions with PPA-optimized code to teach efficiency patterns.
    \textcircled{4}\textbf{Hybrid Architectures:} Combining decomposition, search, and agent-based strategies to manage complexity and computational overhead.
\end{itemize}

\subsection{Copyright}\label{RQ4.3}
Copyright challenges in hardware generation are twofold: protecting the ownership of generated IP and preventing the reproduction of proprietary designs.


\paragraph{(1) Current Methods}
Research addresses these opposing needs via:
\begin{itemize}
    \item \textbf{Ownership Protection:} RTLMarker~\cite{135} embeds watermarks through token and statement-level transformations (e.g., parameter encoding, redundant logic). These signatures survive synthesis, enabling ownership verification at both RTL and gate-level netlists.
    \item \textbf{Infringement Prevention:} FreeV~\cite{61} mitigates reproduction risks through rigorous dataset curation. By applying multi-layer filtering (license screening, deduplication), it significantly reduces infringement rates in fine-tuned models while maintaining generation quality.
\end{itemize}



\paragraph{(2) Challenges \& Future Directions}
Despite these advances, significant copyright alignment challenges persist in LLM-generated Verilog code. 

\begin{itemize}
    \item \textbf{Limitations:} 
    \textcircled{1} Current watermarking techniques primarily focus on ownership protection but struggle to balance robustness against detection transparency; 
    \textcircled{2} Existing copyright infringement detection mechanisms rely heavily on surface-level similarity metrics (e.g., cosine similarity).
    \item \textbf{Solutions:} 
    \textcircled{1}\textbf{Watermarking:} Techniques must survive complex synthesis optimizations while remaining transparent. Future hierarchical schemes could preserve attribution across abstraction levels (behavioral to gate-level).
    \textcircled{2}\textbf{Infringement Detection:} Surface-level similarity metrics often fail to catch functionally equivalent but syntactically distinct copies. Hardware-specific benchmarks incorporating functional similarity and provenance tracking mechanisms are needed to distinguish between standard interface reuse and actual IP theft.
\end{itemize}

\subsection{Hallucinations}\label{RQ4.4}
Hallucinations in LLM-generated Verilog code present a significant alignment challenge, manifesting as code that appears syntactically correct but contains functional errors, logical inconsistencies, or design flaws that only become apparent during testing or implementation. 
Research addressing these challenges has emerged along several promising directions.

\paragraph{(1) Current Methods} Various approaches have been developed to mitigate hallucinations in LLM-generated Verilog code, which can be categorized into three main strategies.
\begin{itemize}
    \item \textbf{Knowledge-Enhanced Generation:} Augmenting models with domain rules reduces errors. HAVEN~\cite{75} employs a three-stage pipeline (interpretation, enhancement, reinforcement) based on a specific error taxonomy. HDLCoRe~\cite{122} uses HDL-aware Chain-of-Thought for self-verification, while ChatModel~\cite{95} leverages multi-agent collaboration to verify designs at functional block levels.
    \item \textbf{Data-Driven refinement:} improving training data quality addresses root causes. RTL++~\cite{66} enriches data with control/data flow graphs to capture structural dependencies. Others focus on error classification~\cite{136} or incorporating explicit reasoning paths into fine-tuning datasets (VeriReason~\cite{85}).
    \item \textbf{Decoding \& Process Optimization:} Modifying generation dynamics improves reliability. DecoRTL~\cite{123} uses syntax-aware temperature adaptation and contrastive decoding to reduce redundancy. Paradigm-based methods~\cite{137} utilize expert-inspired templates for different circuit types to minimize error propagation.
\end{itemize}

\paragraph{(2) Challenges \& Future Directions}
\begin{itemize}
    \item \textbf{Limitations:} 
    \textcircled{1} Hardware design demands absolute functional correctness, making even minor logical errors fatal; 
    \textcircled{2} Current detection methods are often computationally expensive or limited in scope (e.g., missing timing or edge cases).
    \item \textbf{Solutions:} 
    \textcircled{1}\textbf{Verification:} Future work should integrate comprehensive verification (combining formal methods and simulation) directly into generation pipelines.
    \textcircled{2}\textbf{Hybrid Approaches:} Multi-stage approaches that separate structural generation from functional optimization, along with hybrid strategies combining knowledge injection and decoding constraints, offer promising paths to robustly mitigate hallucinations.
\end{itemize}

\begin{table*}[!t]
\centering
\caption{Comparison of alignment approaches for Verilog generation (one row per system).}
\label{tab:alignment_summary}
\resizebox{\textwidth}{!}{
\begin{tabular}{llll}
\toprule
\textbf{System} & \textbf{Year} & \textbf{Key Technique} & \textbf{Alignment Target} \\
\midrule
\multicolumn{4}{l}{\textit{Security --- vulnerability mitigation and threat defense}} \\
SecFSM~\cite{126} & 2025 & KG-guided FSM generation & Prevent unintentional FSM vulnerabilities \\
SecV~\cite{127} & 2025 & Hardware-CWE KG retrieval & Verify security constraints at generation \\
RTL-Breaker~\cite{128} & 2025 & Training-data poisoning study & Assess backdoor attack surface \\
VeriContaminated~\cite{129} & 2025 & Min-K\%/CDD contamination detection & Ensure evaluation integrity \\
SALAD~\cite{130} & 2025 & Machine unlearning & Remove malicious/sensitive patterns \\
\midrule
\multicolumn{4}{l}{\textit{Efficiency --- PPA-aware generation and optimization}} \\
MCTS-RTL~\cite{90} & 2024 & PPA-aware MCTS search & Optimize area-delay product \\
ChipGPT~\cite{70} & 2023 & Enumerative candidate selection & Reduce area/power vs.\ naive LLM output \\
VeriOpt~\cite{131} & 2025 & ICL + multi-role LLM & PPA-aware code refinement \\
Chip Design Agent~\cite{132} & 2025 & Program analysis + discovery search & Reuse-driven PPA improvement \\
HiVeGen~\cite{94} & 2025 & Hierarchical submodule decomposition & Accelerator-level PPA \\
AutoSilicon~\cite{93} & 2025 & Task decomposition + multi-version voting & Industrial-scale PPA \\
VeriPPA~\cite{133} & 2025 & Two-stage correctness $\rightarrow$ PPA & Sequential functional then PPA tuning \\
RTLSquad~\cite{134} & 2025 & Squad-based multi-agent debate & Interpretable PPA trade-offs \\
\midrule
\multicolumn{4}{l}{\textit{Copyright --- ownership protection and infringement prevention}} \\
RTLMarker~\cite{135} & 2025 & RTL/netlist watermarking & Traceable IP ownership \\
FreeV~\cite{61} & 2025 & License-filtered dataset curation & Reduce reproduction of proprietary RTL \\
\midrule
\multicolumn{4}{l}{\textit{Hallucinations --- functional correctness and error mitigation}} \\
HAVEN~\cite{75} & 2025 & Error taxonomy + 3-stage pipeline & Domain-knowledge correction \\
HDLCoRe~\cite{122} & 2025 & HDL-aware CoT + self-verification & Rule-guided generation without tools \\
ChatModel~\cite{95} & 2025 & Multi-agent IR + block verification & Hierarchical design correctness \\
RTL++~\cite{66} & 2025 & CFG/DFG graph enrichment & Structural dependency modeling \\
RTL Error Study~\cite{136} & 2025 & Error classification + remediation rules & Targeted error mitigation \\
VeriReason~\cite{85} & 2025 & Reasoning traces + GRPO & Functional correctness via RL \\
DecoRTL~\cite{123} & 2025 & Syntax-aware temperature + reranking & Decoding-time reliability \\
Paradigm-Based~\cite{137} & 2025 & Expert paradigm templates & Type-specific circuit generation \\
\bottomrule
\end{tabular}}
\end{table*}

Table~\ref{tab:alignment_summary} spans 23 systems across four dimensions. Security work remains attack-heavy; efficiency methods converge on closed-loop PPA feedback; copyright aligns protection (watermarking) with prevention (dataset curation); hallucination mitigation shifts from taxonomy/decoding fixes toward reasoning- and verification-aware training.

\begin{tcolorbox}[left=4pt,right=4pt,top=2pt,bottom=2pt,boxrule=0.5pt]
    \textbf{Answer to RQ4:} 
    Our analysis reveals that alignment of LLM-generated Verilog code requires addressing four key dimensions: security, efficiency, copyright, and hallucinations. 
    Table~\ref{tab:alignment_summary} lists 23 alignment systems across security, efficiency, copyright, and hallucinations, each targeting a distinct alignment objective.
    Current approaches demonstrate promising results but still face significant challenges.
    Future research should focus on developing comprehensive verification frameworks, multi-stage generation pipelines, and hybrid approaches that combine multiple alignment strategies.
\end{tcolorbox}

%% file: sec/08_discussion.tex
\section{The Road Ahead}\label{ChaAndOpp}

In this section, we first synthesize quantitative evidence across reviewed studies, then analyze current limitations, and finally outline a research roadmap for next-generation Verilog generation.

\subsection{Cross-Study Evidence Synthesis}\label{synthesis}

While RQ1--RQ4 provide a taxonomic organization of the field, this subsection offers a structured cross-study comparison to draw evidence-based conclusions.
Since substantial heterogeneity in experimental setups (prompt templates, sampling temperatures, benchmark versions, hardware environments) renders direct comparison of absolute numbers across papers unreliable, we adopt two complementary strategies: (1)~\textit{relative improvement analysis}, extracting within-paper gains under controlled conditions; and (2)~\textit{convergent finding identification}, synthesizing conclusions consistently supported by multiple independent studies.

\subsubsection{Relative Improvement Analysis}

Table~\ref{tab:cross_study} reports within-paper improvements of representative methods over their respective baselines under controlled conditions. Because entries span different benchmarks, cross-row comparisons should be interpreted with caution.

\begin{table*}[!t]
\centering
\caption{Relative improvement ($\Delta$pass@1) over baselines, extracted from controlled within-paper experiments.}
\label{tab:cross_study}
\resizebox{\textwidth}{!}{
\begin{tabular}{llllcl}
\toprule
\textbf{Category} & \textbf{Method} & \textbf{Base Model} & \textbf{Benchmark} & \textbf{$\Delta$pass@1} & \textbf{Baseline} \\
\midrule
\multicolumn{6}{l}{\textit{Training-free Methods}} \\
\midrule
EDA Feedback (Single) & RTLFixer~\cite{139} & GPT-4 & VerilogEval-v1 & +10.1\% & Zero-shot GPT-4 \\
EDA Feedback (Multi) & VerilogCoder~\cite{145} & GPT-4 Turbo & VerilogEval-v2 & +33.9\% & Zero-shot GPT-4 Turbo \\
Prompt Engineering & HDLCoRe~\cite{122} & Qwen2.5-72B & RTLLM-v2 & +10.0\% & Zero-shot Qwen2.5-72B \\
Inference Optimization & VRank~\cite{124} & GPT-4o & VerilogEval-v1 & +9.2\% & Greedy decoding \\
\midrule
\multicolumn{6}{l}{\textit{Supervised Fine-Tuning}} \\
\midrule
Data-Centric & RTLCoder~\cite{80} & DeepSeek-Coder-6.7B & VerilogEval-v1 & +11.4\% & Base DeepSeek-Coder-6.7B \\
Data-Centric & CodeV~\cite{71} & DeepSeek-Coder-6.7B & VerilogEval-v1 & +22.8\% & Base DeepSeek-Coder-6.7B \\
Strategy-Centric & PyraNet~\cite{65} & CodeLlama-7B & VerilogEval-v1 & +3.8\% & Standard SFT \\
Knowledge-Enhanced & VeriThoughts~\cite{87} & Qwen2.5-14B & VerilogEval-v1 & +13.0\% & Base Qwen2.5-14B \\
\midrule
\multicolumn{6}{l}{\textit{Reinforcement Learning}} \\
\midrule
Structure-based & VeriSeek~\cite{86} & DeepSeek-Coder-6.7B & VerilogEval-v1 & +8.3\% & SFT-only \\
Tool-based & VeriPrefer~\cite{84} & DeepSeek-Coder-6.7B & VerilogEval-v1 & +3.1\% & SFT-only \\
\bottomrule
\end{tabular}}
\end{table*}

Three patterns emerge.
(1)~SFT delivers larger gains than prompt engineering or inference optimization among single-model methods, while multi-agent systems (e.g., VerilogCoder) achieve the highest improvements at the expense of increased API costs.
(2)~RL further improves over SFT baselines by 3--8~pp, suggesting that execution-aware optimization complements supervised learning.
(3)~Reasoning distillation yields substantial gains on larger backbones (e.g., VeriThoughts +13.0\% on 14B), offering a qualitatively distinct advantage over data-centric scaling.

\subsubsection{Convergent Findings Across Independent Studies}

We further identify findings that converge across multiple studies despite differing experimental setups:

\noindent\textbf{Finding 1: Domain-adapted small models rival general-purpose large models.}
Multiple studies~\cite{80,71,55,59} independently show that 7B models fine-tuned on Verilog data match or exceed GPT-3.5-turbo and GPT-4 on functional correctness, providing strong evidence that domain specialization compensates for scale.

\noindent\textbf{Finding 2: RL over SFT yields consistent but bounded gains.}
All RL-based studies~\cite{86,84,85,72,83} report 3--8~pp improvements over SFT baselines across different algorithms (PPO, GRPO, DPO), indicating that RL addresses exposure bias and the absence of execution-aware optimization in supervised training.

\noindent\textbf{Finding 3: Multi-agent systems excel on complex designs.}
Comparisons between single- and multi-agent architectures~\cite{145,146,144,93} consistently show gains that scale with design complexity (e.g., VerilogCoder +33.9\% on VerilogEval-v2), suggesting that agent specialization with iterative EDA feedback is most valuable when tasks exceed single-model capacity.

\noindent\textbf{Finding 4: Reasoning distillation is a pivotal 2025 innovation.}
Works leveraging reasoning traces from frontier models (e.g., DeepSeek-R1)~\cite{87,72,68,79} consistently report substantial improvements, with VeriThoughts achieving +13.0\% on a 14B backbone. Unlike data scaling, reasoning distillation introduces structured problem decomposition, addressing a qualitative bottleneck in hardware design understanding rather than merely increasing exposure to Verilog tokens.

\subsubsection{Threats to Comparability}

Several factors limit the above conclusions to directional trends rather than precise effect sizes:
\begin{itemize}[leftmargin=*]
    \item \textbf{Benchmark versioning:} VerilogEval-v1 and v2 differ in difficulty and testbench coverage, precluding cross-version comparison.
    \item \textbf{Prompt sensitivity:} Variations in prompts can shift pass@1 by 5--15 pp even on identical benchmarks~\cite{108}.
    \item \textbf{Evaluation protocol:} Differences in temperature, $n$ in pass@$k$, and post-processing introduce systematic bias.
    \item \textbf{Unreported details:} Many studies omit confidence intervals or generation parameters, precluding formal meta-analysis.
\end{itemize}
Nevertheless, the convergent findings, derived from consistent directional conclusions across heterogeneous settings, offer the community reliable evidence-based guidance on the relative merits of competing paradigms.

\subsection{Limitations}\label{Challenges}

Based on our analysis across RQ1--RQ4 and the synthesis above, we identify four key limitations of current research:

\noindent\textbf{(1) Foundation \& Knowledge Gap.}
General-purpose LLMs lack intrinsic understanding of hardware constraints (timing, concurrency, area). Predominant autoregressive architectures struggle with Verilog's parallel semantics, often producing code that simulates correctly but fails synthesis.
\textit{Future Direction: Develop specialized architectures pre-trained on hardware corpora with integrated EDA feedback.}

\noindent\textbf{(2) Data \& Benchmark Scarcity.}
High-quality Verilog data is scarce compared to software languages, often lacking documentation or testbenches. Existing benchmarks are small-scale ($<$100 samples), focusing on basic modules while neglecting system-level complexities (e.g., pipelined datapaths, SoCs).
\textit{Future Direction: Construct large-scale, uncontaminated benchmarks with graduated complexity and comprehensive verification suites.}

\noindent\textbf{(3) Evaluation \& Alignment Deficits.}
Current metrics (functional-pass@$k$) ignore critical hardware dimensions like PPA and security. Alignment research is nascent: security tools miss hardware-specific vulnerabilities, PPA optimization lacks efficient feedback loops, and copyright mechanisms are fragile.
\textit{Future Direction: Establish standardized frameworks integrating synthesis-based PPA assessment, security analysis, and robust IP protection.}

\noindent\textbf{(4) Deployment Readiness.}
No current solution integrates seamlessly into industrial EDA workflows. The lack of interactive refinement, explainability, and human-in-the-loop features hinders practical adoption by hardware engineers.
\textit{Future Direction: Prioritize IDE-integrated tools supporting iterative collaboration and explainable design decisions.}

\subsection{Roadmap}\label{Roadmap}
Building on the evidence synthesis and identified limitations, we propose a three-stage trajectory to advance the field from isolated module generation toward integrated, system-level solutions.

\subsubsection{Stage 1: Strengthening Foundations}
\begin{itemize}[leftmargin=*]
    \item \textbf{Hardware-Aware Models:} Move beyond general-purpose LLMs by developing billion-scale models pre-trained on hardware-specific corpora. Key innovations should include hardware-aware tokenization and architectural modifications to model concurrency.
    \item \textbf{Comprehensive Benchmarking:} Construct living benchmarks exceeding 1,000 samples, covering the full spectrum from basic gates to industrial SoCs, grounded with synthesis-verified PPA truth.
\end{itemize}

\subsubsection{Stage 2: Expanding Capabilities}
\begin{itemize}[leftmargin=*]
    \item \textbf{System-Level Hierarchical Generation:} Shift focus from single modules to entire systems. This requires automatic decomposition, interface synthesis, and hierarchical planning to manage cross-module dependencies and protocols.
    \item \textbf{Multimodal Inputs:} Enable generation from diverse engineering artifacts (timing diagrams, block schematics, and waveforms), allowing designers to define architectures visually.
\end{itemize}

\subsubsection{Stage 3: Optimization \& Deployment}
\begin{itemize}[leftmargin=*]
    \item \textbf{Closed-Loop PPA \& Security:} Replace open-loop generation with reinforcement learning pipelines driven by direct EDA feedback (synthesis reports, formal verification), ensuring outputs meet physical and security constraints by construction.
    \item \textbf{Human-AI Collaboration:} Develop ``Copilot-for-Hardware'' interfaces that support interactive refinement, explain design choices, and learn from user corrections, facilitating integration into production flows.
\end{itemize}

By synthesizing these directions, the community can transition from experimental prototypes to trustworthy, expert-level assistants that fundamentally transform the electronic design automation ecosystem.

%% file: sec/09_conclusion.tex
\section{Threats to Validity}\label{threats}

\noindent\textbf{Paper search omission.}
Despite searching six databases and employing snowballing to identify 102 relevant papers (including 32 preprints), the fast-paced nature of LLM research may lead to overlooking emerging work. We mitigated this through rigorous keyword selection and multi-stage filtering.

\noindent\textbf{Study selection bias.}
The reduction from initial candidates to the final set involved subjective quality assessments. To address this, we enforced strict inclusion criteria and a standardized Likert-scale framework, requiring a threshold score (12/15) for all entries.

\noindent\textbf{Categorization and analysis bias.}
Categorizing diverse methodologies (models, datasets, alignment strategies) introduces potential classification bias. We minimized this by aligning with established taxonomies and focusing on factual extraction over subjective interpretation, though our roadmap inevitably reflects our synthesized perspective on the field's trajectory.

\section{Conclusion}\label{conclusion}
This systematic literature review analyzes the transformative impact of Large Language Models on Verilog code generation, synthesizing insights from 102 studies (2020–2025).
We comprehensively mapped the landscape of base and instruction-tuned models (RQ1), scrutinized the critical role of curated datasets and evaluation metrics (RQ2), and categorized optimization strategies ranging from prompt engineering to reinforcement learning (RQ3). Furthermore, we highlighted essential alignment dimensions (i.e., security, efficiency, copyright, and hallucinations) critical for practical deployment (RQ4).
By identifying key limitations and proposing a multi-stage research roadmap, this work aims to guide the community in transitioning from experimental prototypes to robust, system-level hardware design assistants, ultimately reshaping the future of Electronic Design Automation.